\documentclass[pdflatex,sn-mathphys-num]{sn-jnl}% Math and Physical Sciences Numbered Reference Style 
\newcommand{\mcs}[1]{\mathcal{S}_{#1}}

\usepackage{graphicx}%
\usepackage{multirow}%
\usepackage{amsmath,amssymb,amsfonts}%
\usepackage{bm}%
\usepackage{bbm}%
\usepackage{amsthm}%
\usepackage{mathrsfs}%
\usepackage[title]{appendix}%
\usepackage{xcolor}%
\usepackage{textcomp}%
\usepackage{manyfoot}%
\usepackage{booktabs}%
\usepackage{algorithm}%
\usepackage{algorithmicx}%
\usepackage{algpseudocode}%
\usepackage{listings}%
\usepackage{tikz}
\usetikzlibrary{shapes,arrows.meta,positioning}%
\usepackage{subcaption}%
\theoremstyle{thmstyleone}%
\theoremstyle{thmstyletwo}%

\theoremstyle{thmstylethree}%

\usepackage{geometry}

\begin{document}

\title[Article Title]{Analysis of a household-scale model for the invasion of Wolbachia into a resident mosquito population}% for control against dengue infection}


\author*[1]{\fnm{Abby} \sur{Barlow}}\email{ahb48@bath.ac.uk}
\author[1]{\fnm{Sarah} \sur{Penington}}\email{sp2355@bath.ac.uk}
\author[1]{\fnm{Ben} \sur{Adams}}\email{ba224@bath.ac.uk}


\affil[1]{\orgdiv{Department of Mathematical Sciences}, \orgname{University of Bath}, \orgaddress{\street{Claverton Down}, \city{Bath}, \postcode{BA2 7AY}, \state{Somerset}, \country{United Kingdom}}}

%\affil[2]{\orgdiv{Department}, \orgname{Organization}, \orgaddress{\street{Street}, \city{City}, \postcode{10587}, \state{State}, \country{Country}}}

%\affil[3]{\orgdiv{Department}, \orgname{Organization}, \orgaddress{\street{Street}, \city{City}, \postcode{610101}, \state{State}, \country{Country}}}


\abstract{In areas infested with Aedes aegypti mosquitoes it may be possible to control dengue, and some other vector-borne diseases, by introducing Wolbachia-infected mosquitoes into the wildtype population. Thus far, empirical and theoretical studies of Wolbachia release have tended to focus on the dynamics at the community scale. However, Ae. aegypti mosquitoes typically dwell in and around the same houses as the people they bite and it can be insightful to explore what happens at the household scale where small population sizes lead to inherently stochastic dynamics. Here we use a continuous-time Markov framework to develop a stochastic household model for small populations of wildtype and Wolbachia-infected mosquitoes. We investigate the transient and long term dynamics of the system, in particular examining the impact of stochasticity on the Wolbachia invasion threshold and bistability between the wildtype-only and Wolbachia-only steady states previously observed in deterministic models. 
We focus on the influence of key parameters which determine the fitness cost of Wolbachia infection and the probability of Wolbachia vertical transmission. Using Markov and matrix population theory, we derive salient characteristics of the system including the probability of successful Wolbachia invasion, the expected time until invasion and the probability that a Wolbachia-infected population reverts to a wildtype population. These attributes can inform strategies for the release of Wolbachia-infected mosquitoes. In addition, we find that releasing the minimum number of Wolbachia-infected mosquitoes required to displace a resident wildtype population according to the deterministic model, only results in that outcome about $20$\% of the time in the stochastic model; a significantly larger release is required to reach a steady state composed entirely of Wolbachia-infected mosquitoes $90$\% of the time.}




\maketitle

\section{Introduction}
Dengue is a vector-borne disease with an estimated global burden of $100$ to $400$ million cases and $40,000$ deaths per year \cite{zeng2021global}. It circulates endemically across tropical and subtropical regions, affecting over $100$ countries \cite{hasan2016dengue,bhatt2013global}, and local transmission has recently been observed in parts of Europe \cite{WHO}. The viral infection is passed to humans via the bite of infected mosquitoes. Its most common vector is the Aedes aegypti mosquito \cite{WHO}. Aedes aegypti is predominantly found in urban and sub-urban areas. It favours artificial water containers, such as unsealed cisterns, tyres and empty plant pots as habitats to lay eggs \cite{ECDC} and has benefited from rapid growth and urbanisation of the human population \cite{trewin2021mark}. \\

Dengue prevention methods typically focus on reducing or eliminating the vector population. Spraying insecticides both indoors and outdoors is often used to suppress the mosquitoes. However, complete elimination requires continual spraying, which can lead to the mosquitoes building up resistance to the insecticides. Dengue vaccines have been developed, but currently have limited scope since they are only advised for individuals who have previously contracted the virus \cite{CDCvac}. An alternative dengue control method that has been actively explored in recent years involves infecting the mosquitoes with Wolbachia. Wolbachia are a group of intracellular bacteria found in arthropods and nematodes \cite{werren2008Wolbachia,hughes2013modelling}. They can alter their host's biology in numerous ways. When present in mosquitoes such as Aedes aegypti, the bacteria can reduce the incidence of vector-borne disease via several mechanisms which reduce the mosquito population size and vector competence \cite{cdc}. Here we will not explore the disease dynamics directly but will focus on the establishment and maintenance of Wolbachia infection in a mosquito population via cytoplasmic incompatibility (CI). CI means that eggs produced by females that are not Wolbachia-infected (hereafter referred to as wildtype) and fertilized by males that are Wolbachia-infected are not viable; eggs produced by Wolbachia-infected females and fertilized by any male are viable, and the majority of offspring are Wolbachia-infected. Therefore, in populations where Wolbachia is present, infected females have a reproductive advantage over uninfected females which, combined with vertical transmission, can drive increasing Wolbachia prevalence with each generation \cite{dorigatti2018using,hughes2013modelling, sinkins2004wolbachia}. The invasion of Wolbachia into a susceptible vector population is, however, not guaranteed. There is a trade-off between the reproductive advantage gained by infected mosquitoes through CI and a fitness cost they experience due to infection. Consequently, when Wolbachia is introduced into a susceptible population, it will only go to fixation (i.e. $100\% $ prevalence) if the initial prevalence exceeds a certain level known as the invasion threshold \cite{dorigatti2018using}. \\

Many cage and field experiments on establishing Wolbachia-infected mosquitoes in wildtype populations have been carried out over the past decade \cite{ross2021designing,dos2022estimating,hoffmann2014invasion,dufault2022disruption}. Mathematical modelling has been used to support and explore the insights generated by these empirical studies. The majority of models neglect spatial structure and assume well-mixed populations  \cite{hughes2013modelling,ndii2012modelling,ndii2015modelling}. Some models describe the movement of mosquitoes through continuous space using dispersal kernels in reaction-diffusion frameworks \cite{turelli2017deploying}; others are highly complex simulations that admit limited analytical tractability \cite{hancock2019predicting}. The dynamics of Wolbachia invasion at the scale of a town or city may be reasonably approximated by deterministic models such as these and formulated with ordinary or partial differential equations. However, Aedes aegypti mosquitoes often live in and around the same dwellings as the people they bite and the invasion dynamics unfold stochastically within small household mosquito populations. Wolbachia releases may even be conducted at the household scale with the objective of generating localised protection against dengue \cite{o2018scaled}. Here, we develop and analyse a stochastic model to investigate the invasion dynamics of Wolbachia-infected mosquitoes at the household scale. We use Markov process theory and matrix population theory to examine how the deterministic invasion threshold, which dictates whether an initial Wolbachia release spreads to the entire population, is characterised in a stochastic setting. We elucidate the medium and long term transient dynamics of the stochastic system to derive key quantities useful in Wolbachia release design, including the probability that Wolbachia-infected mosquitoes successfully invade the household, the expected time until invasion and the probability of reversion to a wildtype population.\\

The structure of the paper is as follows. In Section \ref{sec:mean-field}, we review a deterministic mean-field model of wildtype and Wolbachia-infected mosquito population dynamics \cite{hughes2013modelling} in a household context and discuss the bistability of the steady states where only wildtype or only Wolbachia-infected mosquitoes are present. In Section \ref{sec:stoch}, we recast the model in a stochastic framework and introduce the methods required to analyse the dynamics of Wolbachia invasion and persistence. In Section \ref{sec:3mosq}, we use a tutorial model with very small mosquito populations to elucidate the theory introduced in the previous section. In Section \ref{sec:30mosq}, we extend the model to incorporate more realistic bounds on the mosquito population size and investigate the invasion threshold and bistability in a stochastic setting. Finally, in Section \ref{sec:reversion}, we consider the reversion of a Wolbachia-infected population to wildtype due to partially effective vertical transmission. 

\section{Deterministic mean-field model of wildtype and Wolbachia-infected mosquito dynamics}\label{sec:mean-field}
Here we review a deterministic mean-field model for the dynamics of a population composed of wildtype and Wolbachia-infected mosquitoes. The model was first introduced by \cite{hughes2013modelling}. We consider a slightly simplified version given by
\begin{align}\label{eq:hughes}
\frac{dN_m}{dt}&=b Z_m F(N_m + N_w) - dN_m,\\
 \label{eq:hughes2}   \frac{dN_w}{dt}&=b Z_w F(N_m + N_w) - d' N_w.
\end{align}
Here, $N_m$ is the adult female wildtype mosquito population density and $N_w$ is the adult female Wolbachia-infected mosquito density. As in the original model, male population densities are implicitly assumed to be equal to the female densities. The per capita birth and death rates of the wildtype mosquitoes are denoted $b$ and $d$ respectively and the per capita death rate for the Wolbachia-infected mosquitoes is $d'=\delta d$, where $d\geq 1$. The quantities $Z_m$ and $Z_w$ are given by

\begin{equation}\label{eq:zm}
      Z_m=\frac{N_m}{N_m+N_w}(N_m+(1-v)\phi N_w) + \frac{N_w}{N_m+N_w}((1-u)N_m+(1-v)\phi N_w),
\end{equation}
\begin{equation}\label{eq:zw}
    Z_w=v\phi N_w.
\end{equation}

The $Z_m$ term describes wildtype and Wolbachia-infected females mating with wildtype males, and with Wolbachia-infected males, to produce wildtype offspring. The fitness cost of infection, which acts on reproduction, is $1-\phi$. A proportion $u$ of the offspring of wildtype females and infected males are non-viable due to CI. A proportion $v$ of the offspring of infected females are infected due to vertical transmission. The $Z_w$ term describes Wolbachia-infected females mating with any male to produce infected offspring. Hence $bZ_m$ and $bZ_w$ describe the maximum rates at which wildtype and Wolbachia-infected offspring are produced, in the absence of intraspecific competition, by $N_m$ wildtypes and $N_w$ Wolbachia-infected female mosquitoes. We have slightly revised the expression for $Z_m$ given in \cite{hughes2013modelling} to account for the fact that the offspring when infected males mate with infected females are always viable. 

\begin{figure}
    \centering
    \includegraphics[width=0.5\linewidth]{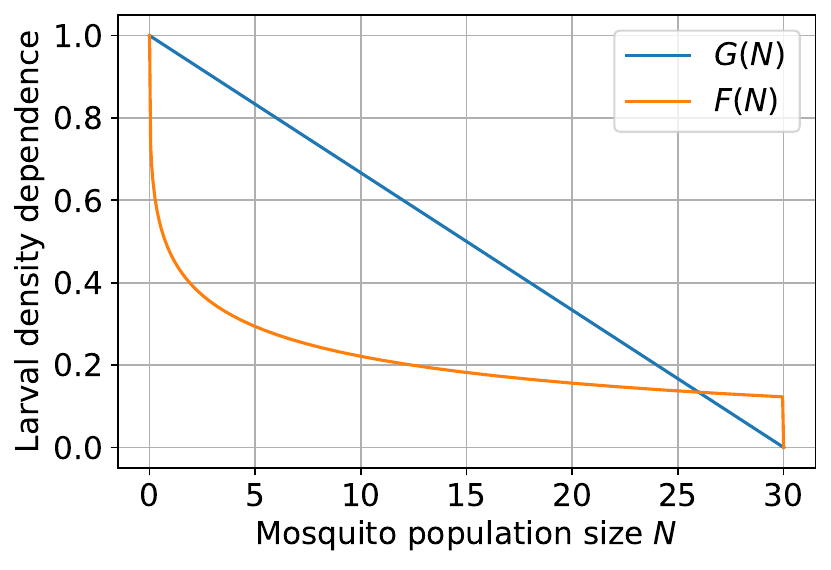}
    \caption{Larval intraspecific competition functions $F(N)$ and $G(N)$; see equations \eqref{eq:hughes_F} and \eqref{eq:qu}. Parameter values are $C=30$, $k=0.3$ and $h=0.76$.}
    \label{fig:larval_density_funcs}
\end{figure}

The function $F(N)$ in equations \eqref{eq:hughes}--\eqref{eq:hughes2} describes intraspecific competition between mosquito larvae. We use the heuristic function \cite{dye1984models}
\begin{align}\label{eq:hughes_F}
    F(N)=\begin{cases}\exp(-hN^k) & N < C,\\
    0 & N \geq C,
    \end{cases}
\end{align}
where $h$ and $k$ describe characteristics of larval competition.
Here we have modified the model in \cite{hughes2013modelling} such that the offspring rate is $0$ when the adult population is $C$ or more. In effect this means that the household mosquito population cannot grow above $C$. Bounding the population size in this way ensures our stochastic model is tractable. Also note that the original model uses $N = Z_m + Z_w$ but, in the interest of parsimony, we use $N = N_m + N_w$.  A simpler alternative to $F(N)$ is $G(N)$ given by \cite{qu2022modeling} 
\begin{align}\label{eq:qu}
    G(N) = \begin{cases} 
    1-N/C & N < C,\\
    0 & N \geq C.
    \end{cases}
\end{align}
Since both functions are heuristic, in Section \ref{ap:dye} in the Appendix, we reproduce our results using $G(N)$ for comparison. Figure \ref{fig:larval_density_funcs} shows the two larval intraspecific competition functions.

Model parameterisation is based on \cite{hughes2013modelling}. We have re-scaled the model such that $N_m$ and $N_w$ are in terms of mosquitoes per $100$m$^2$, the approximate area of a dwelling. We have fixed the reproductive carrying capacity, and effective upper bound on the female mosquito population size, at $C=30$. This bound lies well above observed mosquito household population sizes reported in the literature \cite{madewell2019associations}. All parameter definitions and values are given in Table \ref{tab:params}. 

\begin{center}
\begin{table}[h]
%\begin{minipage}{280pt}
\caption{Parameter definitions and values. Values except $C$ and $b$ are from \cite{hughes2013modelling} and re-scaled so that mosquito population density is expressed per $100$m$^2$. All rates are per day. $C$ is based on \cite{madewell2019associations}, $b$ is calculated such that the wildtype female steady-state in the absence of any Wolbachia infection is $N_m^* = 10$.}\label{tab:params}%
\begin{tabular}{|l|l|l|}
%\toprule
\hline
Parameter& Definition& Value\\
%\midrule
\hline
$b$& Maximum per capita birth rate for wildtype mosquitoes. & $4.52d=0.54$ ($x^*=10$)\\ [10pt]
$d$& Per capita death rate of wildtype mosquitoes.& $0.12$\\ [10pt]
 $b'=\phi b$& \multirow{2}{280pt}{Maximum per capita birth rate for Wolbachia-infected mosquitoes.}& $0.54\phi$\\[25pt]
 $\phi$& \multirow{2}{280pt}{$1-\phi$ is the fitness cost of infection, which acts on reproduction. See $b^\prime$ above.}& $0$--$1$\\[25pt]
$d'=\delta d$& \multirow{2}{280pt}{Per capita death rate of Wolbachia-infected mosquitoes.}& $0.12$ \\[25pt]
 $\delta$& \multirow{2}{280pt}{Mortality cost of infection. See $d^\prime$ above.}&$1$\\[15pt]
$u$& \multirow{2}{280pt}{Cytoplasmic incompatibility. Probability that offspring produced by a Wolbachia-infected male mating with a wildtype female are not viable.}& $1$\\[25pt]
 $v$& \multirow{2}{280pt}{Vertical transmission probability.}&$0.9-1$\\[25pt]
 $C$ & Reproductive carrying capacity. & $30$ \\[10pt]
 $k$& Larval competition parameter.&  $0.3$\\ [10pt]
 $h$& Larval competition parameter.& $0.19\times 100^k=0.76$\\ [10pt]
 \hline
\end{tabular}
%\end{minipage}
\end{table}
\end{center}

\subsection{Stability analysis} \label{subsec:stab_deterministic}
We now summarise the steady state behaviour of the deterministic model \eqref{eq:hughes}--\eqref{eq:hughes2}. There are four steady state solutions: all mosquitoes absent $E_0=(N^*_m,N^*_w)=(0,0)$; only wildtype mosquitoes present $E_1=(F^{-1}(d/b),0)$; only Wolbachia-infected mosquitoes present $E_2=(0,F^{-1}(d'/(bv\phi)))$ and a coexistence state where both wildtype and Wolbachia-infected mosquitoes are present, $E_3$, which can be found numerically. If $b>d$, the absence steady state $E_0$ is unstable and $E_1$ is stable. If in addition $\phi b > \delta d$ then $E_2$ is also stable and the coexistence steady state $E_3$ is a saddle point \cite{hughes2013modelling}.

\begin{figure}
    \centering
    \includegraphics[width=0.5\linewidth]{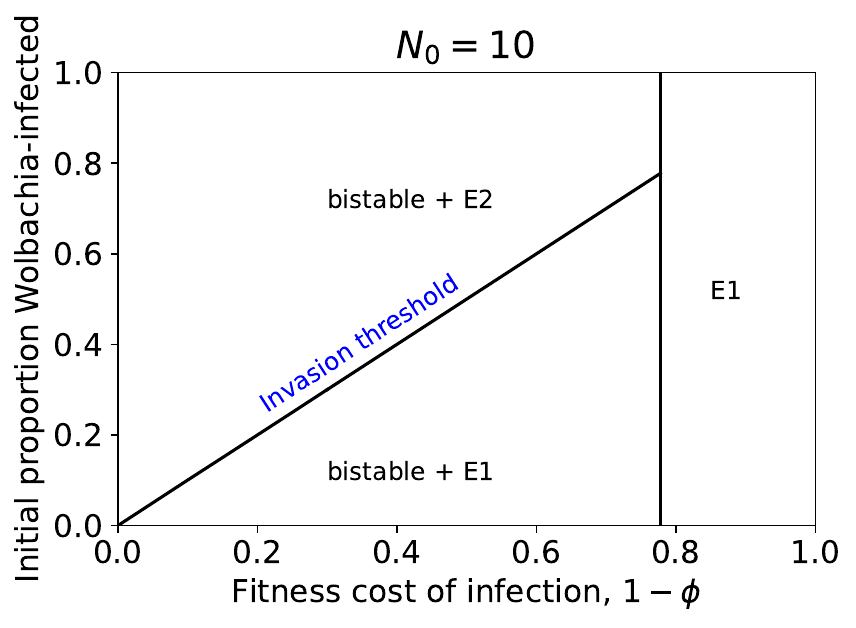}
    \caption{Stability and attraction of the steady states of system \eqref{eq:hughes}--\eqref{eq:hughes2}, with larval density function \eqref{eq:hughes_F}, depending on the fitness cost of Wolbachia infection $1-\phi$ and the initial proportion of the mosquito population that is Wolbachia-infected $N_w(0)/N_0$. Here $u,v=1$ and $N_0 = 10$. Indicated on the diagram are the regions where the system is bistable and either the wildtype-only ($E1$) or Wolbachia-only ($E2$) steady state is attracting, and where the unique non-trivial steady state is wildtype-only, and stable. The invasion threshold under the bistable region is labelled in blue text. All of our stability analysis plots are produced using the open source software bSTAB \cite{stender2022bstab}.}
    \label{fig:basin_attr_Dye}
\end{figure}

Linear stability analysis shows that system \eqref{eq:hughes}-\eqref{eq:hughes2} is bistable in $E_1$ and $E_2$ when CI and vertical transmission are highly efficient i.e. $u$ and $v$ are both close to 1 \cite{hughes2013modelling}. This means that the outcome of introducing Wolbachia-infected mosquitoes into a household depends on the initial numbers of wildtype and infected mosquitoes in the household. Figure \ref{fig:basin_attr_Dye} shows the stability and attraction of the wildtype-only and Wolbachia-only steady states. The proportion of a mosquito population that must be infected in order for the Wolbachia-only state to be attracting defines the invasion threshold. For a household with a total of $N_0=10$ (female) mosquitoes, the invasion threshold in terms of $N_w(0)/N_0$ is shown as a function of the reproductive fitness cost of Wolbachia infection $1-\phi \in [0,1]$. For our parameter values, the steady state (female) wildtype-only population is $10$, so this threshold corresponds to the proportion of the wildtype mosquito population that would need to be replaced with Wolbachia-infected mosquitoes for eventual replacement of the entire wildtype population. When $1-\phi$ is above $1-d/b(=0.78)$, the fitness cost of infection is so high that the Wolbachia-only steady state is not biologically feasible. However, as $1-\phi$ decreases (and the fitness cost of infection reduces) the Wolbachia-only state becomes viable, and the invasion threshold decreases. The bistability is a key feature of the system. It shows that successfully establishing Wolbachia in a mosquito population requires the release of a sufficient number of infected mosquitoes, and that number depends on the fitness cost of infection associated with the Wolbachia strain.  

\begin{figure}
    \centering
    \includegraphics[width=0.5\linewidth]{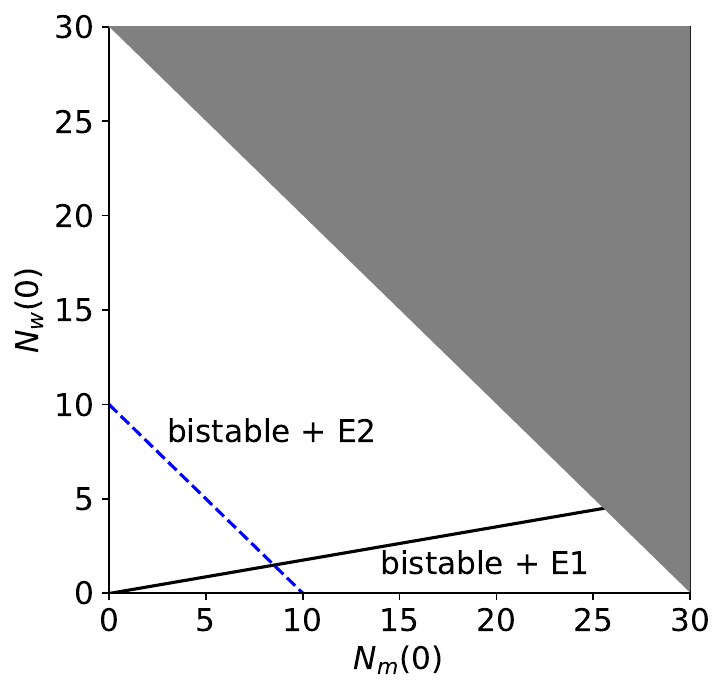}
\caption{Stability and attraction of steady states of system \eqref{eq:hughes}--\eqref{eq:hughes2} depending on the initial numbers of female wildtype $N_m(0)$ and Wolbachia-infected $N_w(0)$ mosquitoes. Here $u,v=1$ and $1-\phi=0.15$, with larval density function \eqref{eq:hughes_F}. Indicated on the diagram are the regions where the wildtype-only ($E1$) or Wolbachia-only ($E2$) steady states are attracting. The grey triangle covers populations outside our upper bound on the total female mosquito population size. The dashed blue line indicates where $N_m(0)+N_w(0)=10$, corresponding to $1-\phi = 0.15$ in Figure \ref{fig:basin_attr_Dye}.}  \label{fig:stab_colour_Dye}
\end{figure}

In Figure  \ref{fig:basin_attr_Dye} we assume the total female population size is fixed at $N_0 = 10$. We now relax this assumption. Figure \ref{fig:stab_colour_Dye} shows the basins of attraction in the bistable region when the initial numbers of (female) wildtype $N_m(0)$ and Wolbachia-infected $N_w(0)$ mosquitoes can take any values consistent with a total population size less than the reproductive carrying capacity ($C = 30$). In this figure, $1-\phi=0.15$ is fixed and the line $N_m(0)+N_w(0)=10$ corresponds to $\phi = 0.85$ in Figure \ref{fig:basin_attr_Dye}. We see that, regardless of the total population size, the Wolbachia-only steady state is achieved if the proportion of Wolbachia-infected mosquitoes exceeds approximately $0.15$.  

\section{The stochastic framework} \label{sec:stoch}
We now present a stochastic formulation of the model. The underlying structure is a continuous-time Markov chain (CTMC) $\mathcal{X}:=\{X(t),t\geq 0\}$ \cite{norris1998markov,anderson2012continuous} which describes the demographic evolution of the wildtype and Wolbachia-infected mosquito populations within a household. While the deterministic model describes the dynamics of the household population sizes directly, the stochastic model describes the dynamics of the probabilities for the household state $X$. The state space is $\{(0,0)\} \cup \mathcal{S}$. Here $\mathcal{S}=\{(m,w) : 0< m+w\leq C\}$ is a finite set of transient states, where $m$ denotes the number of wildtype female mosquitoes and $w$ denotes the number of Wolbachia-infected female mosquitoes in the household. In order to keep the model computationally tractable, we impose an upper bound of $C$ on the number of mosquitoes that can inhabit a household at a given time. State $(0,0)$ corresponds to the absence of both mosquito populations, and is absorbing.

The CTMC describes how the state probabilities of the system evolve over time. A key component of the CTMC is the transition (or generator) matrix $\bm{Q}_{0,0}$, which has the block matrix form \cite{van2013quasi}
\begin{align}
    \bm{Q}_{0,0}=\begin{pmatrix}
        0 & \bm{0} \\
        \bm{a}^T & \bm{Q}
    \end{pmatrix}.
\end{align}
Here, $\bm{Q}=(q_{ij})$ is the transition matrix of the (sub) Markov chain which excludes the absorbing state $(0,0)$. The entry $q_{ij}$ is the rate of transition from state $i$ to state $j$, for $i \ne j$, and $q_{i,i}=-q_i=-\sum_{j\in \mcs{}, \{i\neq j\}\cup \{(0,0)\}}q_{i,j}$. Note that in $\bm{Q}$, the probability dynamics of any state that communicates with the absorbing state, for example $(n,0)$ or $(0,n)$ for some $n \leq C$, still includes the rate of transition to $(0,0)$. In this way, probability leaks out of the system; $\bm{Q}$ does not conserve probability. 

The vector $\bm{a}=(a_i,\;i\in\mathcal{S})$ is composed of the rates of transition to the absorbing state from each state in $\mathcal{S}$. The $\bm{0}$ denotes a row vector of zeros of size $\vert \mathcal{S} \vert$. The dynamics of the household state probabilities are described by a system of linear differential equations, often referred to as the Master, Kolmogorov forward or ensemble equations. Let $\bm{P}_{0,0}(t)$ be the row vector of the probabilities that a household is in each possible state at time $t$, with the first element corresponding to the absorbing state $(0,0)$. Then 
\begin{align}\label{eq:ME}
    \frac{d\bm{P}_{0,0}(t)}{dt}=\bm{P}_{0,0}\bm{Q}_{0,0}.
\end{align}
Solving this equation gives
\begin{align}\label{eq:continuous}
    \bm{P}_{0,0}(t)=\bm{P}_{0,0}(0)\exp{(\bm{Q}_{0,0}t)}.
\end{align}
Alternatively, since $\bm{P}_{0,0}(t)$ is the solution at time $t$ of a set of $s$ linear differential equations, where $s$ is the cardinality of the state space, if the matrix $\bm{Q}_{0,0}$ is diagonalisable, 
\begin{align}\label{eq:cnt_eigs}
    \bm{P}_{0,0}(t)=\sum_{i=1}^s c_ie^{-\lambda_it}\bm{l}_i.
\end{align}
Here, $-\lambda_i$ is the $i^{th}$ eigenvalue of $\bm{Q}_{0,0}$, $\bm{l}_i$ is its corresponding left eigenvector and the $c_i$ are constants \cite{keeling2008methods}. Recall that $\bm{Q}_{0,0}$ has all the eigenvalues of $\bm{Q}$ plus an additional eigenvalue $0$ corresponding to the absorbing state. We can also write the solution as $\bm{P}_{0,0}(t) = \sum_{i=1}^s q_i e^{-\lambda_it}\bm{w}_i$ (for $\bm{Q}_{0,0}$ diagonalisable) where the eigenvalues $-\lambda_i$ are as above, but the constants $q_i$ and right eigenvectors $\bm{w}_i$ are, in general, different. Note that the matrix $\bm{Q}_{0,0}$ may not be diagonalisable, and we will not assume this in the proofs in Appendix \ref{ap:proofs}, but the expression in \eqref{eq:cnt_eigs} is useful for illustrative purposes.

The equation \eqref{eq:ME} can also be decomposed to consider the transient states or absorbing states separately. Let $\bm{P}(t)$ be the state probability vector of the transient subsystem. The solution of $\frac{d\bm{P}(t)}{dt} = \bm{P}(0)\bm{Q}$ using transient state matrix $\bm{Q}$ gives the probability dynamics of the system. Note that the probability mass gradually moves from the transient states to the absorbing state $(0,0)$ which is not included in $\bm{P}(t)$ \cite{van2013quasi}. Thus, the sum of $\bm{P}(t)$ over all states will not generally equal $1$. However, as discussed in \cite{mubayi2019studying} normalising $\bm{P}(t)$ so that it sums to $1$ gives the state probability distribution conditional on non-extinction.

Under this framework, independent of the initial state, the system eventually reaches the absorbing state $(0,0)$ and become trapped there. This is indicated by the fact that all the eigenvalues of $\bm{Q}$ have negative real part. In the eigenvalue formulation of the solution \eqref{eq:cnt_eigs}, the terms with larger magnitude eigenvalues decay more quickly. Eventually, only the term associated with the minimal magnitude eigenvalue is non-negligible. The corresponding eigenvector (once normalised) gives the quasi-stationary distribution (QSD) of the system. In the following subsection, we define the QSD and describe how it is derived for our model. We go on to use the QSD to explore the dynamics of the household state probability distribution in the later part of the transient phase.
 
\subsection{The quasi-stationary distribution} \label{subsec:qsd}
In our stochastic model framework, $\bm{P}_{0,0}(t)$ is a row vector composed of the probabilities of a household occupying each of the possible household states. Since the household will eventually become trapped in the absorbing state $(0,0)$, the limit of this probability distribution as $t\rightarrow \infty$ corresponds to the household being in the extinct state $(0,0)$ with probability $1$. However, we are interested in the long term behaviour of the system after the initial transient stage has passed but before extinction is reached. This behaviour is described, in part, by the QSD \cite{mubayi2019studying}. Essentially, when the QSD is reached, the transient state probabilities of the system remain in a fixed ratio while probability continues to move from the transient states to the absorbing state. Therefore, if the transient state probabilities are conditioned on non-extinction, they are approximately stationary in this phase.

When working with a CTMC, the QSD can often be deduced using classical matrix theory and the transition rates of the process \cite{van2013quasi}. If the transient state transition matrix $\bm{Q}$ is irreducible, it is relatively easy to find the QSD. Let $-\alpha$ be the eigenvalue of $\bm{Q}$ with minimal magnitude. Since $\bm{Q}$ is an irreducible transition matrix, $-\alpha$ will be unique, simple and negative \cite{van2013quasi}. It is often referred to as the decay parameter of the Markov chain $\mathcal{X}$. Note that, throughout this paper, minimal magnitude refers to taking the minimal magnitude of the real part of the eigenvalue if it is complex. The QSD is the left eigenvector $\bm{l}$ associated with the eigenvalue $-\alpha$, normalised so that the elements sum to $1$. The decay rate of the probability mass not already in the absorbing state is $\alpha$, and the limiting distribution of that probability, conditional on non-absorption, is the QSD. 

\subsubsection{Reducible Markov chain} \label{subsec:reducible}
When the transition matrix $Q$ is reducible, finding the QSD is more challenging. This is the case with our model. In order to find the QSD when the transition matrix is reducible, the state space must first be partitioned into its communicating classes. Here we follow the approach in \cite{van2013quasi}. For two states $i$ and $j$, we write $i \rightarrow j$ if the probability of reaching state $j$ from state $i$ via a series of state transitions is non-zero. Two states $i$ and $j$ are said to communicate, $i \leftrightarrow j$, if $i \rightarrow j$ and $j\rightarrow i$. The relation $\leftrightarrow$ is an equivalence relation on $\mcs{}$ that partitions $\mcs{}$ into subsets known as communicating classes. If $Q$ is irreducible then the state space is composed of a single communicating class.

Suppose $Q$ is irreducible and $\mcs{}$ is composed of $L > 1$ communicating classes denoted $\mcs{1},\mcs{2}, \ldots , \mathcal{S}_L$. Let $\bm{Q}_k$ be the submatrix of $\bm{Q}$ composed of the transition rates between the states in $\mcs{k}$. A class $\mcs{i}$ is accessible from a class $\mathcal{S}_j$, written $\mcs{i} \prec \mathcal{S}_j$, if starting from any state in $\mathcal{S}_j$, we are able to reach any state in $\mcs{i}$ via some sequence of state transitions. Since $\mcs{i}$ and $\mathcal{S}_j$ are distinct communicating classes, the converse is not true; $\mathcal{S}_j$ cannot be reached from $\mcs{i}$. We order the communicating classes in such a way that $\mcs{i}\prec \mcs{j}$ implies that $j > i$; such an ordering exists because there cannot exist $i,j,k$ distinct with $\mcs{i} \prec \mcs{j} \prec \mcs{k}$ and $\mcs{i}\succ \mcs{k}$, since $\mcs{i}, \mcs{j}, \mcs{k}$ are distinct communicating classes. One might visualise this structure as a cascade --- probability mass always moves through the communicating classes down the ordering. The lowest class in the ordering, $\mcs{1}$, is termed the minimal class \cite{van2013quasi}. The states themselves must be labelled in such a way that $\bm{Q}$ is of lower-triangular block form and the sub-$Q$ matrices, $\bm{Q}_k$, reside along the diagonal of $\bm{Q}$. If necessary, the states can be relabelled to achieve this. The matrices $\bm{Q}_k$ must be ordered as described above, with the matrix $Q_1$ positioned in the top left hand corner of $\bm{Q}$. This is the transition matrix that corresponds to communicating class $\mcs{1}$, which is accessible from at least one of the classes higher up the order e.g. $\mcs{2}$, $\mcs{3}$ etc.

The set of eigenvalues of $\bm{Q}$ is equal to the union of the sets of eigenvalues of the $\bm{Q}_k$ matrices. The minimal magnitude eigenvalues of each $\bm{Q}_k$, which we denote $-\alpha_k$, are all unique (within the set of eigenvalues of $\bm{Q}_k$), simple and negative. Hence, the minimal magnitude eigenvalue of $\bm{Q}$ is the smallest of the set of minimal magnitude eigenvalues of the $\bm{Q}_k$. That is, the minimal magnitude eigenvalue of $Q$ is $-\alpha := -\min_k \alpha_k$ \cite{seneta2006non}. 
The QSD is, as in the irreducible case, the left eigenvector of $Q$ corresponding to $-\alpha$ normalised to sum to $1$. If the algebraic multiplicity of $-\alpha$ is greater than one, then $\alpha = \alpha_k$ in at least two values of $k$. In this case, the QSD is given by the eigenvalue-eigenvector pair of $\bm{Q}$ corresponding to the minimal communicating class, as defined above. 

\subsubsection{Interpretation}\label{subsec:interp}
The matrices $\bm{Q}$ and $\bm{Q}_k$ for $k = 1, \dots, L$ describe the dynamics of probability mass moving between the household states. The eigenvalues also tell us about what is happening in the communicating classes.
In addition to all the eigenvalues of $\bm{Q}$ having negative real part, all of the eigenvalues of each $\bm{Q}_k$ have negative real part. This indicates that the probability mass in communicating class $\mcs{k}$ is decaying to $0$ and the probability is moving to accessible classes further down the order, that is to classes $\mcs{j}$ with $j<k$ or to the absorbing state $(0,0)$.
In each class $\mcs{k}$, the contributions to the probability dynamics associated with the larger magnitude eigenvalues drop away most quickly, and eventually the decay rate of the total probability mass contained within the class is well approximated by the minimal magnitude eigenvalue of the class. We show that this is true for our system and others with the same communicating class structure in Section \ref{ap:claim1} in the Appendix. 
In general, if the minimal magnitude eigenvalue of a class is much more negative than those of the other classes, the dynamics in that class resolve more quickly. The probability mass in that class becomes negligible, accumulating in the other classes.  

In the system as whole, the class with the eigenvalue of smallest overall magnitude holds significant probability mass for longer than the others and thus dominates the QSD. The left eigenvector of $\bm{Q}$ associated with this eigenvalue gives the stable ratio of state probabilities in the class, and indeed the entire system, in the QSD phase. When there is only a small difference in the minimal magnitude eigenvalues between classes, the QSD can still be found mathematically. In practice, however, emergence of the QSD requires all of the probability dynamics to be well approximated by a single eigenvalue term. If the first and second minimal magnitude eigenvalues are similar, most of the probability mass reaches the absorbing state before the single eigenvalue approximation becomes valid. In \cite{keeling2008methods} the authors suggest the difference between the first and second most minimal magnitude eigenvalues needs to be larger than the magnitude of the first in order for the QSD to be meaningful.

\subsection{Communicating class structure of our model}\label{subsec:com-class}
We define our state space $\{(0,0)\} \cup \mathcal{S}$ to be made up of the absorbing state $(0,0)$ (no mosquitoes present) and the finite set of transient states $\mathcal{S}=\{(m,w) : 0< m+w\leq C\}$. As in the mean-field model in Section \ref{sec:mean-field}, $m$ and $w$ record the number of female wildtype and Wolbachia-infected mosquitoes and we implicitly assume there is always an equal number of male mosquitoes. Also in alignment with the mean-field model, we assume that wildtype mosquitoes die at per capita rate $d$, Wolbachia-infected mosquitoes die at rate $\delta d$, wildtype mosquitoes are born at total rate $mbZ_mF(m+w)$ and Wolbachia-infected mosquitoes are born at total rate $wbZ_wF(m+w)$, where $Z_m$ and $Z_w$ are as in equations \eqref{eq:zm}--\eqref{eq:zw}. We interpret one unit of time as one day. The density dependent competition function $F$ is configured so that $F(m+w) = 0$ if $m+w \geq C$, effectively bounding the population size at $C$. Table \ref{tab1} summarises the transition rates for the model.

\begin{center}
\begin{table}[h]
\caption{Transition events and rates from household state $k$ to state $l$ in the CMTC model.}\label{tab1}
\begin{tabular}{|l|l|l|l|}
\hline
State $k$ & State $l$  & Transition & Rate\\
\hline
$(m,w)$ & $(m+1,w)$  & wildtype birth  & $mbZ_mF(m+w)$\\[10pt]
$(m,w)$ & $(m,w+1)$  & Wolbachia-infected birth& $wbZ_wF(m+w)$ \\[10pt]
$(m,w)$ & $(m-1,w)$  & wildtype death & $dm$ \\[10pt]
$(m,w)$ & $(m,w-1)$  & Wolbachia-infected death& $d\delta w$ \\
\hline
\end{tabular}
\end{table}
\end{center}

In the interest of parsimony we do not allow movement between households, thus considering a single isolated household. We assume that the offspring of Wolbachia-infected mosquitoes are always Wolbachia-infected (perfect vertical transmission, $v = 1$). Under this framework, $\mcs{}$ consists of $3$ communicating classes: $\mcs{1}=\{(m,0):0 < m\leq C\}$ is the class of wildtype-only states, $\mcs{2}=\{(0,w):0< w \leq C\}$ is the class of Wolbachia-only states and $\mcs{3}=\{(m,w):m,w > 0 \;\; \text{and} \;\; m+w\leq C\}$ is the class of mixed states. The mixed state class must be labelled as $\mcs{3}$ since the two other communicating classes are accessible from $\mcs{3}$ but $\mcs{3}$ can not be accessed from either of $\mcs{1}$ or $\mcs{2}$.

It is straightforward to construct our $\bm{Q}_k$ matrices, for $k = 1, 2, 3$, and hence $\bm{Q}$ itself, in the required form. The ordering of the states within each sub-$Q$ matrix is inconsequential. The full $\bm{Q}$ matrix has the form
\begin{align}
    \bm{Q}=\begin{pmatrix}
        \bm{Q}_1 & \bm{0} & \bm{0} \\
        \bm{Q}_{21} & \bm{Q}_2 & \bm{0} \\
        \bm{Q}_{31} & \bm{Q}_{32} & \bm{Q}_3
    \end{pmatrix},
\end{align}
where $\bm{Q}_{ij}$ denotes the block matrix containing transition rates from the states in communicating class $\mcs{i}$ to states in the communicating class $\mathcal{S}_j$. Under the current model framework, vertical transmission is perfect ($v = 1$) so there is no way to return from the Wolbachia-only state to a wildtype-only state (via some mixed state) and $\bm{Q}_{21}$ is a matrix of zeros. Recall that the diagonal elements of $\bm{Q}_k$ are not equal to the negative sum of the rest of the corresponding row of $\bm{Q}_k$. Rather, $\bm{Q}_k[i,i]$ is equal to $-\sum_{j \in \mathcal{S}\cup \{(0,0)\}} q_{i,j}$, where $i$ corresponds to a given state in $\mcs{k}$ and $q_{i,j}$ denotes the transition rate from state $i$ to state $j$.

In Section \ref{subsec:interp} we noted that if a communicating class has a much larger minimal magnitude eigenvalue than the other classes, then the probability mass in that class decays quickly in comparison to the other classes. We will see this happening with class $\mcs{3}$ in our model. Since the absorbing state in our model can only be reached through $\mcs{1}$ (wildtype-only states) or $\mcs{2}$ (Wolbachia-only states), the probability mass drains out of $\mcs{3}$ (mixed states) initially, then out of $\mcs{1}$ or $\mcs{2}$ to the absorbing state. Since, when $v=1$, $\mcs{1}$ and $\mcs{2}$ are not communicating, once the system enters $\mcs{1}$ or $\mcs{2}$ it remains in that class until extinction. The time to extinction is governed by the minimal magnitude eigenvalue of the class. Prior to extinction, the state probabilities in the class reach a stable distribution given by the left eigenvector associated with the minimal magnitude eigenvalue. For the mathematical details, see Section \ref{ap:claim2} in the Appendix. 

\subsection{The probability of Wolbachia invasion}\label{subsec:absorbprob}
In Section \ref{subsec:stab_deterministic} we saw that, in the deterministic model, there is a bistability between the wildtype-only and Wolbachia-only steady states. If an initial release causes the proportion of mosquitoes that are Wolbachia-infected to exceed the invasion threshold then, in the deterministic framework, a Wolbachia-infected population permanently replaces the wildtype population. We define a release that results in a household mosquito population becoming entirely Wolbachia-infected as `successful'. In the stochastic framework, exceeding the invasion threshold does not guarantee a successful release, and no mosquito population persists indefinitely. For any initial mixed state, the stochastic process eventually enters the absorbing state $(0,0)$. Before that, it must pass through either the Wolbachia-only class or the wildtype-only class, with probabilities governed by the initial household state and model parameters. So a release that passes through the Wolbachia-only class is successful. 

Here, we will briefly discuss how to utilise standard Markov chain theory \cite{norris1998markov} to calculate the probability of successful Wolbachia invasion. 
As before, let the wildtype-only, Wolbachia-only and mixed type classes be $\mcs{1}$, $\mcs{2}$ and $\mcs{3}$ respectively. We denote the first time at which the Wolbachia-only class is entered as $T=\inf \{t\geq 0 :X(t)\in \mcs{2}\}$, so that the probability of entering into the Wolbachia-only class at state $(0,w')\in \mcs{2}$ having started in state $(m,w)\in \mathcal{S}_3$ is \cite{gomez2012extinction}
\begin{align}
    a_{(m,w)}(0,w')=\mathbb{P}(T<\infty, X(T)=(0,w')|X(0)=(m,w)).
\end{align}
Clearly, the probability of entering the Wolbachia-only class at $(0,w')$ is $1$ if the current state is 
$(0,w')$ and $0$ if it is contained in $\mcs{1}$, $\mcs{2}\setminus \{(0,w')\}$ or if it is $(0,0)$. 

Let $\bm{a}(0,w')=\left(a_{(m,w)}(0,w'): (m,w) \in \mcs{3} \right)$ be a column vector of the probabilities of entering into $\mcs{2}$ at $(0,w')$ from each state in $\mcs{3}$. Let $\bm{q}(0,w')$ be a column vector containing the transition rates from each state in $\mcs{3}$ to $(0,w')$. Then $\bm{a}(0,w')$ can be found by solving the following linear system of equations:
\begin{align}\label{eq:prob_wolb}
    -\bm{Q}_3 \bm{a}(0,w')=\bm{q}(0,w').
\end{align}
Here $\bm{Q}_3$ denotes the sub-$Q$ matrix corresponding to the mixed household states in $\mcs{3}$. The system of equations \eqref{eq:prob_wolb} follows directly from Theorem 3.3.1 in \cite{norris1998markov}, where the subset to be `hit' (reached) is simply $\{(0,w')\}$. The same method can be used analogously to find the probabilities of entering the wildtype-only class at a given state from any given state in $\mcs{3}$. The probability of a successful Wolbachia invasion starting from a mixed state $(m,w)$ is hence the sum of all the individual probabilities of entering $\mcs{2}$ from $(m,w)\in \mcs{3}$ via each of the states in $\mcs{2}$.

Recall that our state space is bounded such that $\mathcal{S}=\{(m,w): 0< m+w\leq C\}$. So the probability of reaching a Wolbachia-only state $(0,K)$ such that $K\geq C$ from any mixed state $(m,w)\in\mcs{3}$ is $0$. This is because $(0,C)$ is reached from the mixed state $(1,C)$ via wildtype mortality, but $(1,C)$ is not a permissible state because the total population size is bounded by $C$.

\subsection{Time until Wolbachia invasion}\label{subsec:absorbtime}
Standard Markov chain theory \cite{norris1998markov} can be extended to find the expected time until Wolbachia invasion. This is the time at which the household first enters a state in the Wolbachia-only class. 

Let $\tau^*_{(k,l)}$ be the expected time until the Wolbachia-only class $\mcs{2}$ is reached from state $(k,l)\in \mcs{2}\cup\mcs{3}$ (a Wolbachia-only or mixed state), conditional on the event that $\mcs{2}$ is reached from state $(k,l)$. Let $a^*_{(k,l)}$ be the probability of reaching $\mcs{2}$ from state $(k,l)$. Once again, we determine the expected time until successful Wolbachia invasion originating from a mixed state in $\mcs{3}$ by solving the following linear system \cite{gomez2012extinction}:
\begin{align}\label{eq:absrobtime}
    \sum_{(k,l)\in \mcs{2} \cup \mcs{3}} q_{(m,w),(k,l)}a^*_{(k,l)}\tau_{(k,l)}^*=-a^*_{(m,w)}\;\;\; (m,w)\in \mcs{3}.
\end{align}
Recall that $\tau^*_{(k,l)}=0$ for any $(k,l) \in \mcs{2}$. That is, the time until a Wolbachia-only state is reached is $0$ if the current household state is Wolbachia-only. Details of the derivation of equations \eqref{eq:absrobtime} are given in the Appendix \ref{ap:absorb_time}.

\subsection{Quantifying the transient dynamics}\label{subsec:trans_dyn_quant}
Here we introduce some tools from matrix population theory \cite{caswell2000matrix} to calculate key quantities in our CTMC model. These quantities give indications of the expected time until the system reaches the QSD and the uncertainty associated with different regions of the state space.

We can derive the damping ratio $\rho$ from equation \eqref{eq:cnt_eigs} as described in \cite{caswell2000matrix}; see Appendix \ref{ap:damping_ratio}. Since we do not include the absorbing state in $\bm{P}(t)$, in our system the damping ratio quantifies the rate of convergence to the QSD and is defined as
\begin{align}
    \rho=\exp((-\lambda_1)-(-\lambda_2)).
\end{align}
The difference between the two eigenvalues is called the spectral gap and is often used to quantify the convergence of a Markov chain \cite{keeling2008methods,dambrine1981note}.
When $t$ is large,
\begin{align}
    \left \vert \left \vert \frac{\bm{P}(t)}{e^{-\lambda_1 t}} - c_1 \bm{l}_1 \right \vert \right \vert &\leq k \rho ^{-t} = ke^{-t\log \rho}.
\end{align}
This means that for some constant vector $k$, convergence to the QSD (denoted $\bm{l}_1$ here) is asymptotically exponential, at a rate at least as fast as $\log \rho$ \cite{caswell2000matrix}. Further, the time $t_x$ at which the contribution of the eigenvalue $-\lambda_1$ to the probability mass becomes $x$ times as great as the contribution of eigenvalue $-\lambda_2$, is found by solving
\begin{align}
    \rho^{t_x} = x \Rightarrow t_x=\log{(x)}/\log{(\rho)}.
\end{align}
We can therefore use the damping ratio as a indicator of how close the system is to the QSD.

The entropy is another useful characterisation of the transient dynamics. Entropy measures how uncertain we are of the next state to be reached given we know the current state. For continuous-time processes \cite{spencer2005continuous} we calculate the entropy using the jump matrix of the CTMC which is composed of the probabilities $p_{i,j}$ that a process in state $i$ moves next to state $j$ 
\begin{align}
     p_{i,j}=\begin{cases}
       -q_{i,j}/q_i & \text{if $i \neq j$ and $q_i\neq0$},\\
       0 & \text{if $q_i=0$},
    \end{cases}
\end{align}
and 
\begin{align}
    p_{i,i} = \begin{cases}
    0 & \text{if $q_i \neq 0$}, \\
    1 & \text{if $q_i = 0$}.
    \end{cases}
\end{align}
Recall that $q_i$ is equivalent to $-q_{i,i}$ from Section \ref{sec:stoch}.
The entropy of state $i$ is defined as
\begin{align}
    H_i= \sum_{j\in \mcs{} \cup \{(0,0)\}} p_{i,j}\log{(p_{i,j})}.
\end{align}
We will use the entropy to identify household states for which the dynamics are most uncertain because they give rise to large numbers of possible stochastic pathways. 

\section{Tutorial model - household with up to 3 female mosquitoes}\label{sec:3mosq}
To aid understanding of the theory outlined above, in this section we work through a tutorial model. We set the upper bound on the number of (female) mosquitoes in the household at $C=3$. These mosquitoes can be wildtype or Wolbachia-infected. This configuration yields a small state space with just $9$ transient states plus $(0,0)$, which facilitates visualisation of the communicating classes, the QSD and the associated dynamics. Figure \ref{fig:schematic1} shows a state transition diagram which details the composition of the communicating classes outlined in Section \ref{subsec:com-class}. Although the maximum household size is much smaller than we would expect to observe in reality, the characteristic dynamics of the system are informative.
\begin{figure}
\centering
\begin{tikzpicture}[ -> , >= Stealth, shorten >=2pt , line width =0.5 pt, node distance =2 cm]
\node[circle, draw, fill=orange] (onezero) at (0,0) {(1,0)};
\node[circle, draw, fill=orange] (twozero) at (2,0) {(2,0)};
\node[circle, draw, fill=orange] (threezero) at (4,0) {(3,0)};
\path (onezero) edge [bend left] node [above] {} (twozero);
\path (twozero) edge [bend left] node [below] {} (onezero);
\path (twozero) edge [bend left] node [above] {} (threezero);
\path (threezero) edge [bend left] node [below] {} (twozero);
\node[circle, draw, fill=gray] (zerozero) at (1,-2) {(0,0)};
\path (onezero) edge [bend left] node [above] {} (zerozero);
\node[circle, draw, fill=blue] (zeroone)  at (0,-4) {\textcolor{white}{(0,1)}};
\path (zeroone) edge [bend right] node [above] {} (zerozero);
\node[circle, draw, fill=blue] (zerotwo) at (2,-4) {\textcolor{white}{(0,2)}};
\path (zeroone) edge [bend left] node [above] {} (zerotwo);
\path (zerotwo) edge [bend left] node [below] {} (zeroone);
\node[circle, draw, fill=blue] (zerothree) at (4,-4) {\textcolor{white}{(0,3)}};
\path (zerotwo) edge [bend left] node [above] {} (zerothree);
\path (zerothree) edge [bend left] node [below] {} (zerotwo);
\node[circle, draw, fill=green] (oneone) at (-1,-2) {(1,1)};
\path (oneone) edge [bend left] node [above] {} (onezero);
\path (oneone) edge [bend right] node [above] {} (zeroone);
\node[circle, draw, fill=green] (twoone) at (-3,0) {(2,1)};
\path (twoone) edge [bend right] node [above] {} (oneone);
\path (oneone) edge [bend right] node [above] {} (twoone);
\path (twoone) edge [bend left] node [below] {} (twozero.north);
\node[circle, draw, fill=green] (onetwo) at (-3,-4) {(1,2)};
\path (onetwo) edge [bend right] node [above] {} (oneone);
\path (oneone) edge [bend right] node [above] {} (onetwo);
\path (onetwo) edge [bend right] node [below] {} (zerotwo.south);
\end{tikzpicture}
\caption{Schematic of the communicating classes in the $3$ mosquito model. The states in the mixed class $\mcs{3}$ are shaded green, the states in the wildtype-only class $\mcs{1}$ are orange and the states in the Wolbachia-only class $\mcs{2}$ are blue.}
\label{fig:schematic1}
\end{figure}

\subsection{Parameterisation}\label{subsec:parameterisation}
We parameterise the $3$ mosquito model as in Table \ref{tab:params} except for setting $C=3$ and $b=0.3$ corresponding to a wildtype-only deterministic steady state of $x^*=2$, and setting the fitness cost of infection $1-\phi = 0.3$. We chose these parameters to effectively illustrate how we use the mathematical techniques and interpret the transient dynamics of the system. In later sections, we will move towards a more realistic parameter regime supported by the literature \cite{hughes2013modelling}. Larval competition parameters $h$ and $k$ remain the same as in Table \ref{tab:params}. No reproduction occurs in households where the population has reached the upper bound. %The per capita birth rate $b$ is set such that the number of mosquitoes in the household at the wildtype-only deterministic steady state is $m^* = x^*$. We chose $x^*=2$, since if $x^*=3$, the per capita birth rate would be undefined.

\subsection{Results}
We use the methodology described in Section \ref{sec:stoch} to determine the QSD and related metrics for the system with households of up to $3$ mosquitoes. Table \ref{tab:3mosq_results} shows the minimal magnitude eigenvalues of each of the communicating classes along with the quasi-stationary distribution derived from the eigenvector associated with overall minimal magnitude eigenvalue. This eigenvalue arises from the wildtype-only class. The second smallest eigenvalue in magnitude arises from the Wolbachia-only class. This indicates that the probability mass leaves $\mcs{2}$ more quickly than it leaves $\mcs{1}$. Consequently, for sufficiently large time, almost all of the probability mass outside of the absorbing state is in $\mcs{1}$, and distributed according to the QSD. The QSD only holds non-zero probability mass in the states of the wildtype-only class. 
The minimal magnitude eigenvalue of class $\mcs{3}$ is much larger than those of the other two classes. This indicates that the probability mass quickly moves out of the mixed class $\mcs{3}$, then more slowly moves out of the wildtype-only and Wolbachia-only classes, $\mcs{1}$ and $\mcs{2}$, and into the absorbing state. 

\begin{figure}
  \centering
  \begin{subfigure}{0.45\textwidth}
    \includegraphics[width=\linewidth]{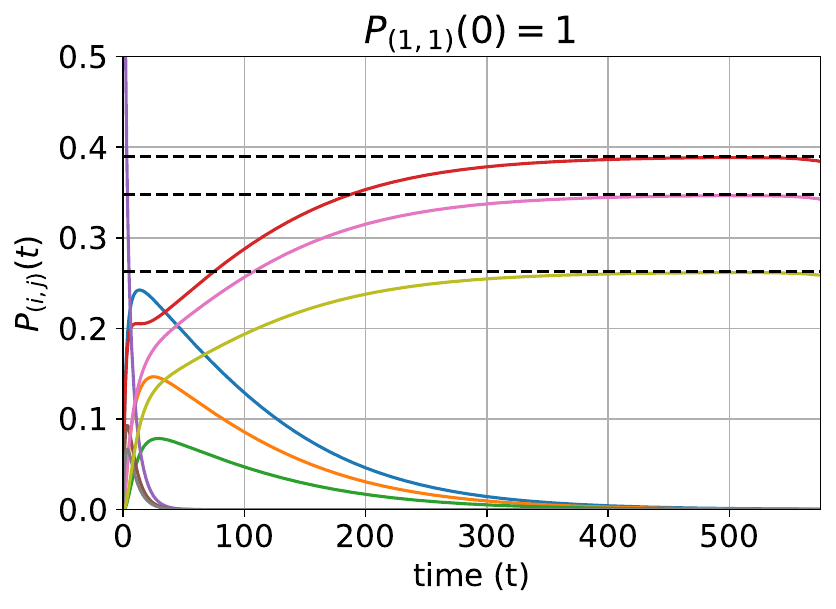}
    \caption{}
    \label{fig:subfig1}
  \end{subfigure}
  \hfill
  \begin{subfigure}{0.45\textwidth}
    \includegraphics[width=\linewidth]{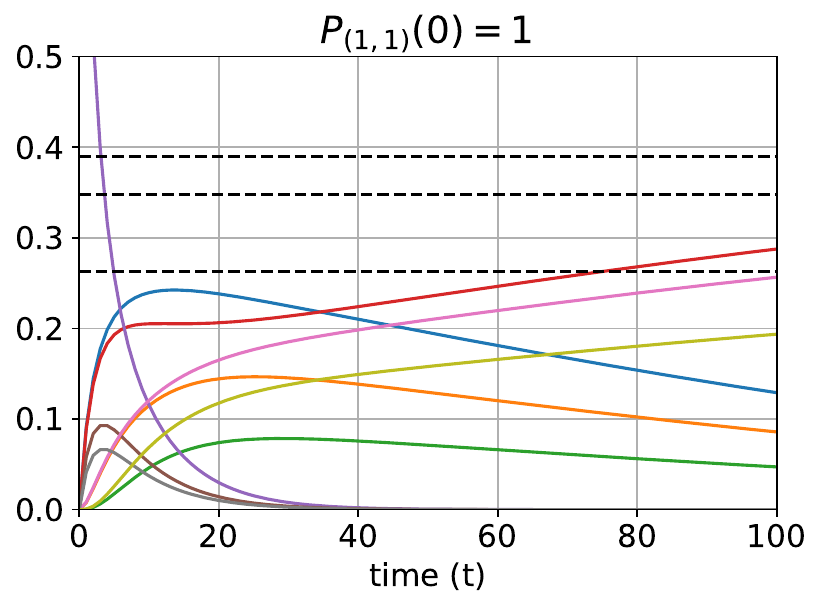}
    \caption{}
    \label{fig:subfig2}
  \end{subfigure}
  \caption{Probability distribution of all transient states over time, conditioned on non-extinction and initial state being $(1,1)$ with probability $1$. Parameter values are as stated in Subsection \ref{subsec:parameterisation}. (a) Over $575$ days and (b) enlargement of the first $100$ days. The dashed lines indicate the QSD values for the wildtype-only states.}
  \label{fig:3mosq}
\end{figure}

In Figures \ref{fig:subfig1} and \ref{fig:subfig2} we show the dynamics of the transient state probability distribution over an interval of $575$ days. We condition on non-extinction and an initial probability distribution $P_{(1,1)}(0)=1$ corresponding to one wildtype and one Wolbachia-infected (female) mosquito. The initial phase of the dynamics is most clear in Figure \ref{fig:subfig2}, which shows the first $100$ days. The probability in the mixed states of class $\mcs{3}$ quickly decays away in the first $40$ days, accumulating in the wildtype-only or Wolbachia-only states of the $\mcs{1}$ and $\mcs{2}$ classes. In the next phase, shown in Figure \ref{fig:subfig1}, the probability mass decays out of the these classes and into the absorbing state. However, the decay is faster from $\mcs{2}$ and the distributions are conditioned on non-extinction. Therefore, the conditioned probability mass in the Wolbachia-only states decays towards zero, and the probability mass in the wildtype-only states steadily increases towards to the QSD. 

The QSD does not depend on the initial number of wildtype and Wolbachia-infected mosquitoes, so cannot be used to determine whether a release of Wolbachia-infected mosquitoes will successfully invade. The QSD is based on the probability mass dynamics of large ensembles of concurrent and independent realisations of the process, in our case corresponding to independent households. Any single realisation of the stochastic process will at some point either enter the wildtype-only class, and remain there until extinction, or enter the Wolbachia-only class and remain there until extinction. Within either of these classes, the probability mass decays to zero at a rate given by the minimal magnitude eigenvalue of the class and with stable state distribution given by the associated eigenvector (see Appendix Sections \ref{ap:claim1} and \ref{ap:claim2} for justification). 

\begin{table}[]
    \centering
    \begin{tabular}{|l|c|} \hline 
Result&Numerical values\\ 
\hline 
\multirow{2}{200pt}{Minimal magnitude eigenvalues for each communicating class}& $-0.047$ $(\mcs{1})$, $-0.052$ $(\mcs{2})$, $-0.183$ $(\mcs{3})$\\[15pt] 
%\hline 
Quasi-stationary distribution & $[0.390, 0.348,0.263,0,0,0,0,0,0]$\\ 
%\hline 
\multirow{2}{200pt}{Probability of reaching Wolbachia-only class from mixed state $(m,w)$}&$0.524$ $(1,1)$, $0.682$ $(1,2)$, $0.349$ $(2,1)$\\[15pt]
%\hline 
\multirow{2}{200pt}{Conditional expected time (days) to reach Wolbachia-only class from mixed state $(m,w)$}&$5.03$ $(1,1)$, $5.35$ $(1,2)$, $7.80$ $(2,1)$\\[15pt]
%\hline 
\multirow{2}{200pt}{Probability of reaching wildtype-only class from mixed state $(m,w)$}&$0.477$ $(1,1)$, $0.318$ $(1,2)$, $0.651$ $(2,1)$\\[15pt]
%\hline 
\multirow{2}{200pt}{Conditional expected time (days) to reach wildtype-only class from mixed state $(m,w)$}&$4.81$ $(1,1)$, $7.59$ $(1,2)$, $5.13$ $(2,1)$\\[15pt]
\hline
\end{tabular}
    \caption{Key results for the model with households of up to $3$ mosquitoes. Minimal magnitude eigenvalues of each class, the probability of reaching the Wolbachia-only class, expected time to reach the Wolbachia-only class conditional on reaching the class, probability and expected time (conditional on reaching the class) to reach the wildtype-only class. In the QSD vector, the first three terms correspond to the states $(1,0)$, $(2,0)$, $(3,0)$.}
    \label{tab:3mosq_results}
\end{table}

Table \ref{tab:3mosq_results} also gives the probabilities of reaching the Wolbachia-only class from each mixed state $(m,w)$ with $m >0$ and $w >0$. We define this to be a successful invasion in the stochastic model context. The complements of these values are the probabilities of reaching the wildtype-only class. From $(1,1)$, the probability of reaching the wildtype-only, and the probability of reaching the Wolbachia-only class, are both close to $0.5$. This is intuitive since the sequences of birth-death events required to reach either class are the same, and the probabilities of these sequences are similar when $\phi$ is close to $1$. If the initial number of wildtype mosquitoes is larger than the number of Wolbachia-infected mosquitoes, the chance of reaching the wildtype class increases because the probability of wildtype births occurring is much larger than that of Wolbachia-infected births. The pattern is reversed if there are initially more Wolbachia-infected mosquitoes. The expected time required to reach the Wolbachia-only class from a mixed state is smallest from $(1,1)$ and largest from $(2,1)$. The expected time required to reach the wildtype-only class is also smallest from $(1,1)$ and largest from $(1,2)$. 

The damping ratio calculated from the minimal eigenvalues of classes $\mcs{1}$ and $\mcs{2}$ allows us to estimate the convergence time to the QSD. The QSD is reached when the remaining probability mass in $\mcs{2}$ is negligible compared to that in $\mcs{1}$. The time until the total probability mass in $\mcs{1}$ is $10$ times that in $\mcs{2}$ is $\log{(10)/\log{(\rho)}}$. This value, along with the expected times for the probability mass in $\mcs{1}$ to reach $100$ and $1000$ times that in $\mcs{2}$ are shown in Table \ref{tab:trans_results}. Notice how these values are much larger, of the order several hundred days, than the expected time of around $6$ days it takes to leave the mixed class $\mcs{3}$ (shown in Table \ref{tab:3mosq_results}).  

\begin{figure}
    \centering
    \includegraphics[width=0.6\linewidth]{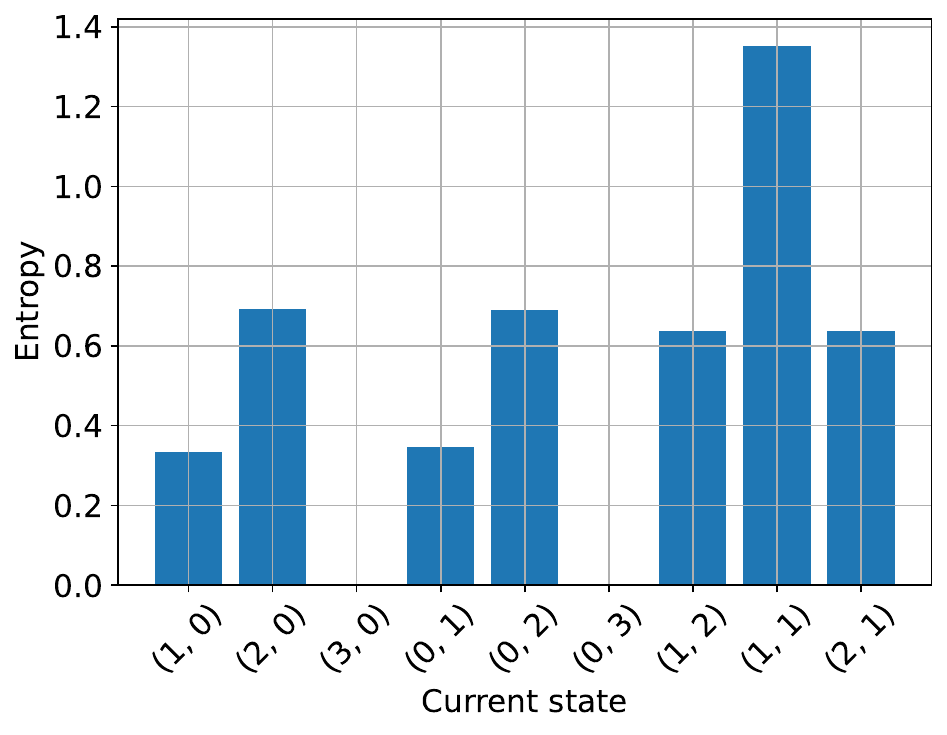}
    \caption{Individual entropies (uncertainty) of the next state to be reached given the current state. Evaluated in the model with households of up to $3$ mosquitoes and parameter values as stated in Subsection \ref{subsec:parameterisation}.}
    \label{fig:entropy_3mosq}
\end{figure}

The individual state entropies further elucidate the transient dynamics of the system. These quantities indicate how certain we are of the next state to be reached given the current state. In Figure \ref{fig:entropy_3mosq} values closer to $0$ correspond to less variable dynamics whereas values closer to $1$ indicate greater variability \cite{brice2020moderate}. There is no variability in the next state that is reached from $(0,3)$ and $(3,0)$ because the only possible event in each case is mortality. The highest variability arises from state $(1,1)$ because there are four possible birth/death events and the probabilities of birth/death events are similar in the wildtype and Wolbachia-infected populations. 

\section{Full model - households of up to $30$ female mosquitoes}\label{sec:30mosq}
We now reparameterise the model so that a household can contain up to $30$ female mosquitoes. This is a more realistic upper bound for a typical infestation. We choose parameters in such a way that in the deterministic model, the wildtype-only steady state is given by $x^*=10$. This value agrees with the mean-field model \eqref{eq:hughes}--\eqref{eq:hughes2} from Section \ref{sec:mean-field} scaled to household area. As before, we set the per capita birth rate $b$ to achieve this steady state. Other parameters are in line with \cite{hughes2013modelling} and given in Table \ref{tab:params}. The fitness cost of Wolbachia infection $1-\phi$ is set to $0.15$ unless otherwise stated. The vertical transmission probability is set to $v=1$, so that Wolbachia-infected mosquitoes can only produce infected offspring and there is no possibility of reversion from a Wolbachia-only state to a mixed state. The total transient state space $\mathcal{S}$ is decomposed as described in Section \ref{subsec:com-class}, where $\mcs{1}=\{(m,0):0 < m\leq 30\}$ is the wildtype-only class; $\mcs{2}=\{(0,w):0< w \leq 30\}$ is the Wolbachia-only class and $\mcs{3}=\{(m,w):m,w > 0 \;\; \text{and} \;\; m+w\leq 30\}$ is the mixed class.

A key aim of our analysis is to explore how stochasticity affects the transient and asymptotic dynamics previously identified in the deterministic system. In particular, we will examine how the deterministic bistability between the wildtype and Wolbachia-only steady states manifests in the stochastic model. 

\subsection{Results}\label{subsec:30mosq_results}
\begin{figure}
    \centering
    \includegraphics[width=0.5\linewidth]{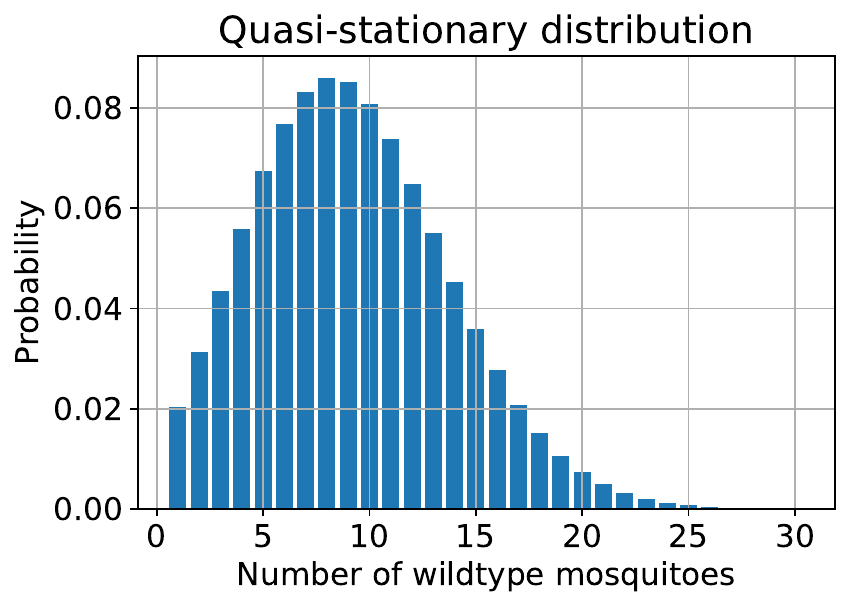}
    \caption{QSD for the model with households of up to $30$ female mosquitoes, perfect vertical transmission ($v=1$), $1-\phi=0.15$ and larval density function \ref{eq:hughes_F}. Any states not shown have probability equal to $0$.}
    \label{fig:qsd_Dye}
\end{figure}

The minimal magnitude eigenvalues of each communicating class are shown in Table \ref{tab:eigs}. As before, the overall minimal magnitude eigenvalue is associated with the wildtype-only class and the second minimal magnitude eigenvalue is associated with the Wolbachia-only class. The difference between these two eigenvalues is larger than the magnitude of the wildtype-only minimal magnitude eigenvalue, implying the QSD is valid. The QSD is shown in Figure \ref{fig:qsd_Dye}. The distribution only has non-zero probability mass in the wildtype-only states and is concentrated around the deterministic steady state $x^*=10$. The probabilities of household states with $28$ or more female mosquitoes are close to $0$, justifying our upper bound on household size. The damping ratio between classes $\mcs{1}$ and $\mcs{2}$ as the system approaches the QSD are shown in Table \ref{tab:trans_results}. In comparison with the model with households of up to $3$ female mosquitoes (Section \ref{sec:3mosq}), the expected number of mosquitoes in the household is much larger (approximately $8$ in the wildtype-only QSD, $5$ in the Wolbachia-only QSD; see Appendix \ref{ap:wolb_QSD}) but the damping ratio indicates that the time to reach the QSD is only $2.4$ times greater.

\begin{table}
\begin{tabular}{|l|c|c|c|}
\hline
    Model & $\mcs{1}$ & $\mcs{2}$ & $\mcs{3}$\\
    \hline
     No reversion & $-0.002$& $-0.007$& $-0.079$\\
     Reversion & $-0.002$& $-0.013$& \\
     \hline
\end{tabular}
\caption{Minimal magnitude eigenvalues of each communicating class in model with up to $30$ female mosquitoes in a household, with ($v=0.9$) and without reversion ($v=1$).}
    \label{tab:eigs}
\end{table}

\begin{table}
\begin{tabular}{|l|c|c|c|}
\hline
     Quantity & $3$ mosquitoes & \multirow{2}{100pt}{\centering $30$ mosquitoes no reversion} & $30$ mosquitoes reversion \\[12pt]
     \hline
    Damping ratio $\rho$ & $1.01$& $1.00$& $1.01$\\
    $\log(10)/\log(\rho)$ & $410$& $977$& $219$\\
    $\log(100)/\log(\rho)$ & $821$& $1954$ & $438$\\
    $\log(1000)/\log(\rho)$ & $1231$ & $2931$ & $656$\\
    \hline
\end{tabular}
\caption{Damping ratio between classes $\mcs{1}$ and $\mcs{2}$, $\rho = \exp(\alpha_1 - \alpha_2)$ and the time until the probability mass in class $\mcs{1}$ is $10$, $100$ and $1000$ times the probability mass in $\mcs{2}$ as the system approaches the QSD. Results are shown for the model with households of up to $3$ female mosquitoes, up to $30$ female mosquitoes with perfect vertical transmission (no reversion) and with imperfect vertical transmission (reversion).}
    \label{tab:trans_results}
\end{table} 

\begin{figure}
    \centering
    \includegraphics[width=0.5\linewidth]{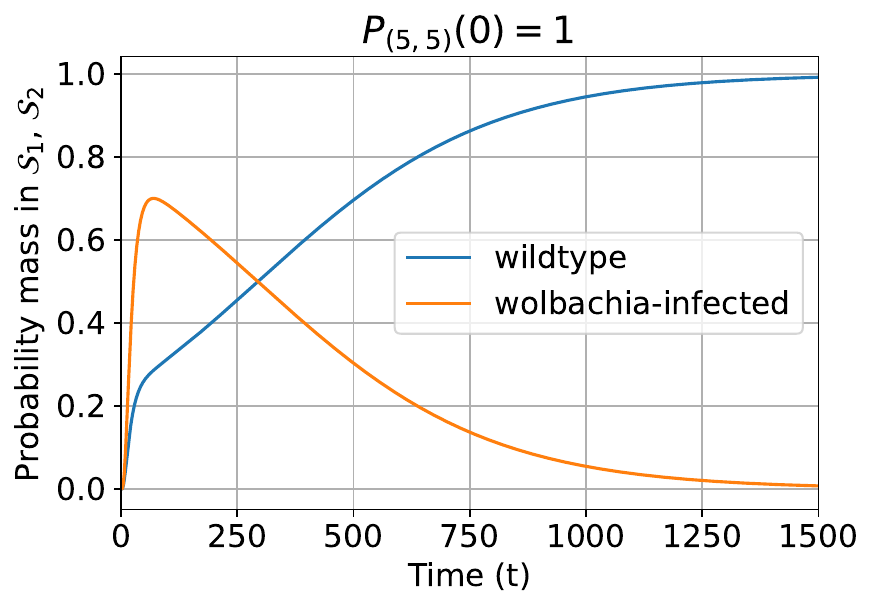}
    \caption{Dynamics of the probability distribution for the wildtype-only and Wolbachia-only classes conditioned on non-extinction. The class probability is the sum of the probabilities for all states in that class. For the model with households of up to $30$ mosquitoes, perfect vertical transmission ($v=1$), $1-\phi=0.15$ and larval density function \eqref{eq:hughes_F}. Initially $P_{(5,5)}(0)=1$. For small $t$ some of the probability mass is in the mixed class $\mcs{3}$ but this quickly decays and has been omitted here.}
    \label{fig:prob_cond_Dye}
\end{figure}

\begin{figure}
    \centering
    \includegraphics[width=0.5\linewidth]{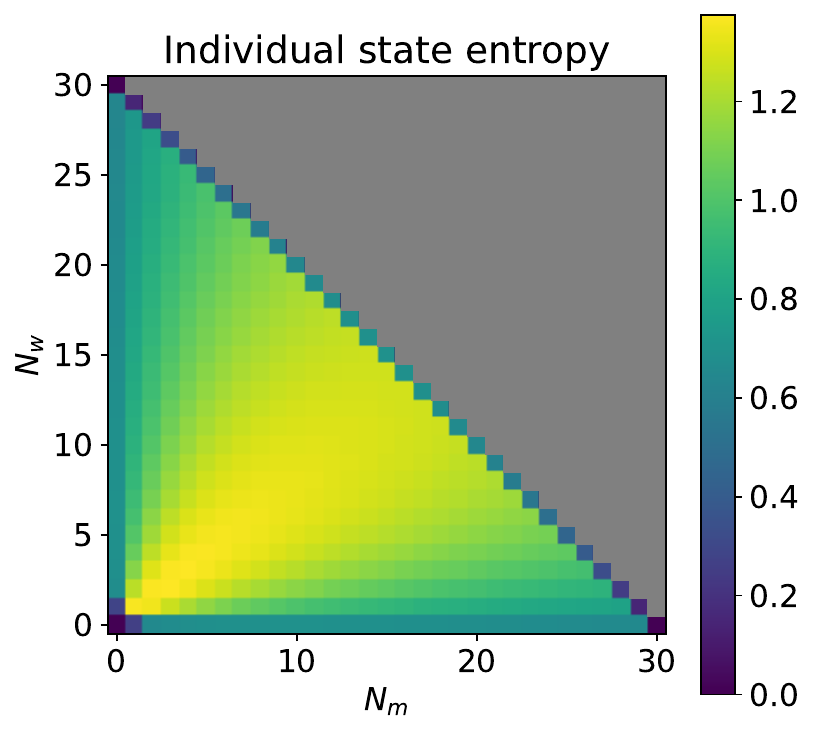}
    \caption{Entropy of state $(N_m,N_w)$. Higher entropy corresponds to greater uncertainty in the next state. The maximum number of female mosquitoes in household is $30$, parameters are as in Table \ref{tab:params}, $1-\phi=0.15$, $v=1$, larval density function is \eqref{eq:hughes_F}. The top right quadrant shaded grey is not an admissible region of the state space.}
    \label{fig:entropies_hughes_Dye}
\end{figure}

Figure \ref{fig:prob_cond_Dye} shows the probability distribution over time for classes $\mcs{1}$ and $\mcs{2}$, conditioned on non-extinction. The initial household state is $(5,5)$ with probability $1$. We chose $(5,5)$ as the initial condition since it has one of the largest individual state entropies (see Figure \ref{fig:entropies_hughes_Dye}) and so allows a broad array of possible stochastic trajectories. The mixed states do not hold significant probability mass for long and are disregarded (not shown in Figure \ref{fig:prob_cond_Dye}). The conditional probability mass in $\mcs{2}$ initially grows faster than in $\mcs{1}$ due to the reproductive advantage of Wolbachia infection conferred by CI. Each class then settles into a regime of exponential decay (to the absorbing state) with a stable state probability distribution well approximated by the QSD for the class in isolation. We provide some analytic insight into these dynamics in Appendix \ref{ap:claim2}. Conditioned on non-extinction as in the Figure \ref{fig:prob_cond_Dye}, the probability accumulates in the wildtype-only class as the fitness cost of infection drives extinction more quickly in the Wolbachia-only class.

\begin{figure}
    \centering
    \includegraphics[width=0.7\linewidth]{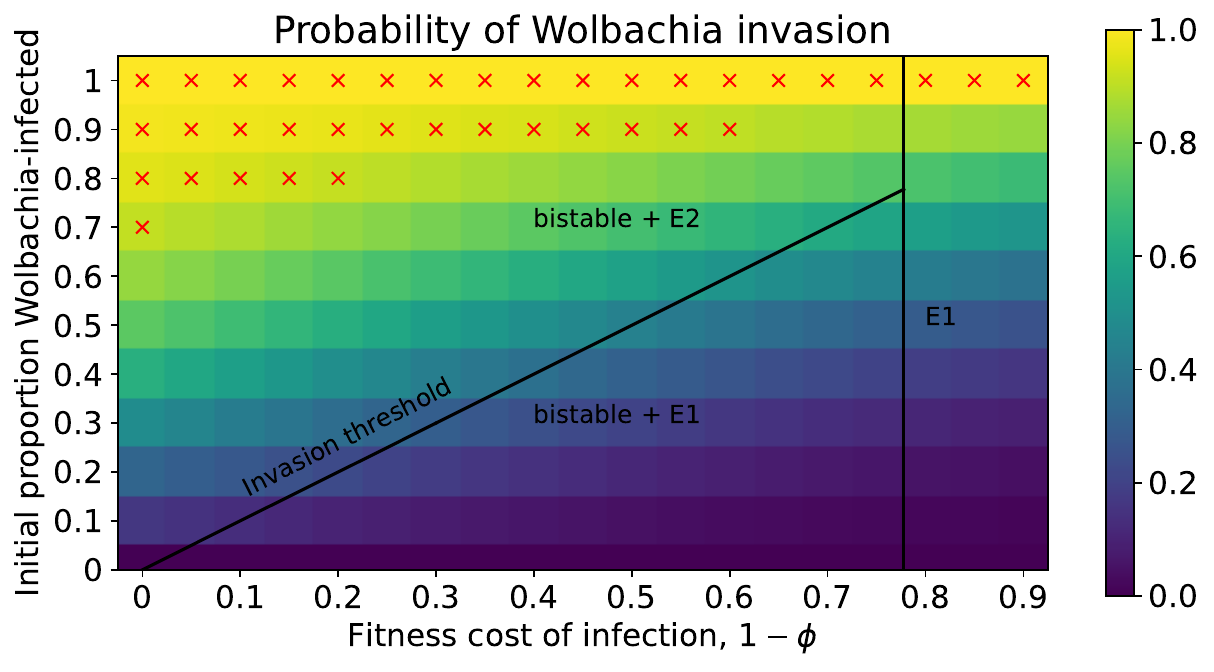}
    \caption{Probability of successful Wolbachia invasion of a household depending on the initial Wolbachia-infected proportion $N_w/N_0$ and fitness cost of infection $1-\phi$ with larval density function \eqref{eq:hughes_F}. The initial household size is fixed at $N_0=10$ female mosquitoes and the initial infected proportion is determined by $N_w(0)$ with $N_m(0) + N_w(0) = N_0$. Households can have up to $30$ female mosquitoes and $v=1$. The solid lines and text show the invasion threshold and stable steady states of the corresponding deterministic model. The red crosses indicate where the probability of successful Wolbachia invasion is greater than or equal to $0.9$.}
    \label{fig:color_stab_dye_v1}
\end{figure}

Figure \ref{fig:color_stab_dye_v1} demonstrates how the deterministic bistability manifests in the stochastic model. Under the deterministic framework, we plotted the invasion threshold for Wolbachia-infected mosquitoes with household size $N_0=10$. This threshold separates the basins of attraction of the wildtype-only and Wolbachia-only steady states and is also shown in Figure \ref{fig:color_stab_dye_v1} for comparison. In the stochastic context, there is no such hard boundary. A Wolbachia invasion is `successful' if the associated stochastic trajectory enters the Wolbachia-only class. Figure \ref{fig:color_stab_dye_v1} shows the probabilities of successful Wolbachia invasions depending on the fitness cost of infection $1-\phi\in[0,0.9]$ and the initial proportion of Wolbachia-infected mosquitoes in the household. If the household contains a non-zero proportion of Wolbachia-infected mosquitoes there is a positive probability of successful Wobachia invasion. This probability decreases as the fitness cost of infection $1-\phi$ increases, and increases as the initial proportion of mosquitoes that are Wolbachia-infected increases. However, the probability of invasion is not constant along the deterministic invasion threshold. When the fitness cost of infection is low, the deterministic threshold suggests that only a small number of Wolbachia-infected mosquitoes are required for successful invasion. However, in the stochastic model, small Wolbachia-infected populations are prone to chance extinction and the probability of successful invasion from the initial population specified by the deterministic threshold is relatively low. The red crosses in Figure \ref{fig:color_stab_dye_v1} indicate where the probability of a successful Wolbachia invasion is greater than or equal to $0.9$. Observe that when $1-\phi=0.15$, at least $80\%$ of the mosquitoes in the household need to be Wolbachia-infected in order for invasion to be successful with probability greater than or equal to $0.9$.
In Figure \ref{fig:color_stab_dye_v1} the total number of mosquitoes in the household is conserved while the proportion of Wolbachia-infected mosquitoes is varied. In Appendix \ref{ap:pop_conserve} we show corresponding results when the household size is not conserved and the introduction of Wolbachia-infected mosquitoes may push the population size above the wildtype-only steady state. 

\begin{figure}
  \centering
  \begin{subfigure}{0.4\textwidth}
    \includegraphics[width=\linewidth]{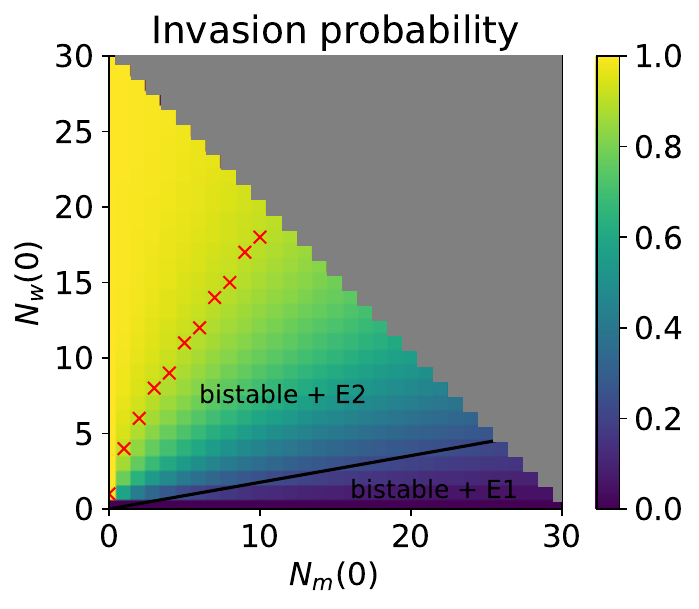}
    \caption{}
    \label{fig:subfig1_prob_invade}
  \end{subfigure}
  \hfill
  \begin{subfigure}{0.4\textwidth}
    \includegraphics[width=\linewidth]{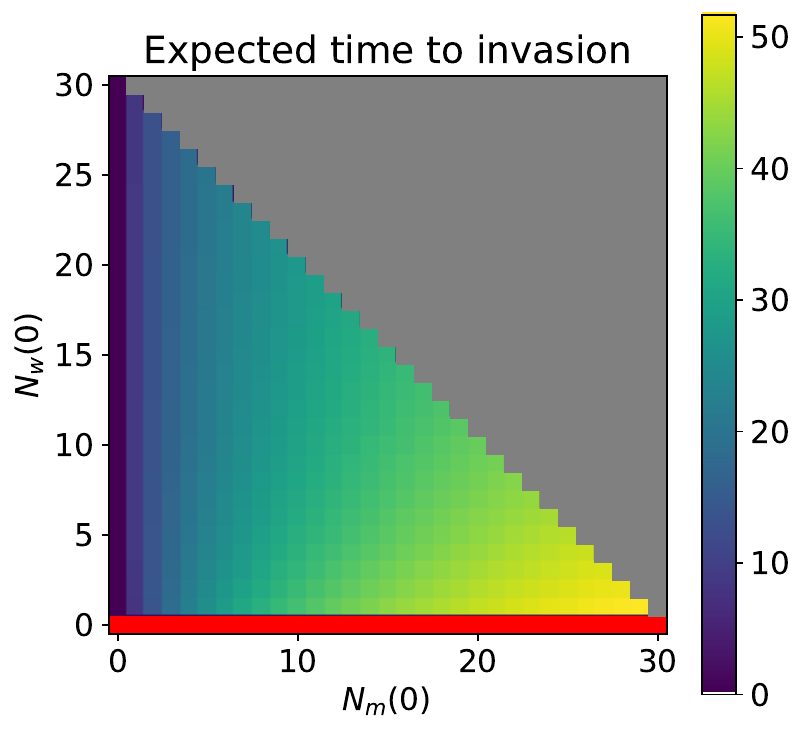}
    \caption{}
    \label{fig:subfig2_time_invade}
  \end{subfigure}
  \caption{(a) Probability of entering the Wolbachia-infected class when the initial state is $(N_m(0), N_w(0))$. The solid black line is the corresponding deterministic threshold (Figure \ref{fig:stab_colour_Dye}). Red crosses denote, for different wildtype mosquito numbers, the minimum number of Wolbachia-infected mosquitoes required for successful invasion with probability $0.9$ or higher. Note that, if there are more than $10$ wildtype female mosquitoes present, the number of Wolbachia-infected mosquitoes required for a successful invasion with probability $0.9$ or higher is outside of the state space. (b) Expected time in days to enter the Wolbachia-infected class, conditional on this event occurring, when the initial state is $(N_m(0), N_w(0))$. The red bar indicates initial conditions where only wildtype mosquitoes are present and so the Wolbachia-infected class cannot be reached. In both diagrams the maximum number of female mosquitoes in a household is $30$, the initial state probability distribution is $P_{(N_m(0),N_w(0))}(0)=1$, parameters are as in Table \ref{tab:params}, $1-\phi=0.15$, $v=1$, and the larval density function is \eqref{eq:hughes_F}. The top right quadrant shaded grey is not an admissible region of the state space.}\label{fig:color_plots_dye_v1}
  \end{figure}

Figure \ref{fig:subfig1_prob_invade} shows how the probabilities of entering the Wolbachia-infected class depend on the initial state $(m,w)$, which allows for resident populations above or below the steady state. As in Figure \ref{fig:color_stab_dye_v1}, we overlay the corresponding deterministic boundary. Along the boundary line, we find that the probability of successful Wolbachia invasion stays fairly consistent; as long as there are at least $3$ wildtype (female) mosquitoes in the household, the probability of invasion is close to $0.23$. The red crosses indicate, for different wildtype mosquito numbers, the minimum number of Wolbachia-infected mosquitoes required for Wolbachia invasion to be successful with a probability greater than or equal to $0.9$. Figure \ref{fig:subfig2_time_invade} shows how the expected time until Wolbachia invasion depends on the initial state. Larger wildtype populations increase the expected time required to enter the Wolbachia-infected class because it takes longer for them to die out. But, in all cases, the time for the invasion dynamics to resolve is of the order $50$ days or less. 

Figure \ref{fig:entropies_hughes_Dye} shows the state entropy for each transient state in $\mathcal{S}$. Larger entropy values correspond to a higher uncertainty in the next state the household will visit. Entropy is highest for the evenly mixed states in $\mcs{3}$ because the number of possible events is maximised, and the event probabilities are evenly weighted. Entropy is lowest for states in the wildtype-only or Wolbachia-only classes because the number of possible events is lower. So the entropy may give an indication of how close the household is to the wildtype-only or Wolbachia-only class. 

\begin{figure}
    \centering
    \includegraphics[width=0.5\linewidth]{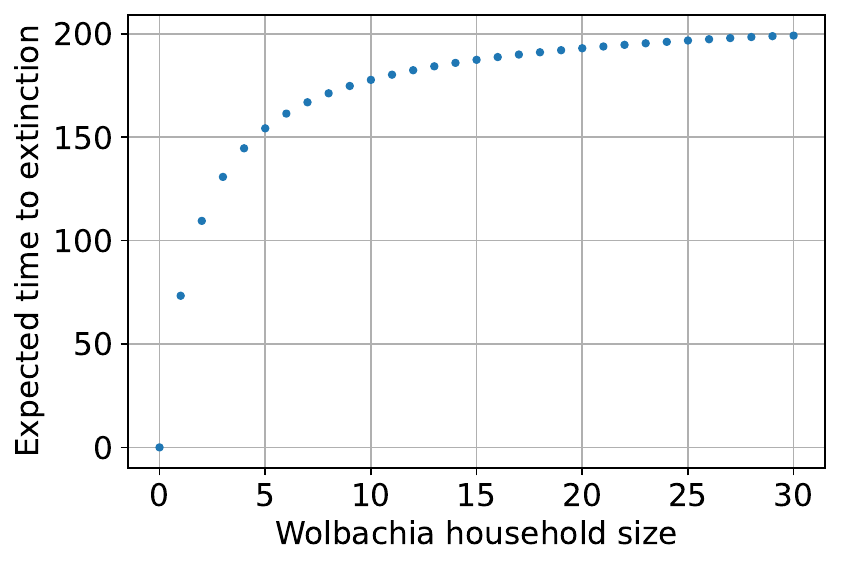}
    \caption{Expected time until extinction for a household population with no wildtype mosquitoes and up to $30$ Wolbachia-infected female mosquitoes. Parameters are as in Table \ref{tab:params}. The fitness cost of Wolbachia infection is $1-\phi=0.15$, $v=1$ and larval density function is \eqref{eq:hughes_F}.}
    \label{fig:ext_times}
\end{figure}

In contrast to the deterministic model, in the stochastic framework a successful Wolbachia invasion does not establish the population permanently. The small population size and random walk of the population dynamics eventually lead to extinction and the household enters the absorbing state $(0,0)$. It is then vulnerable to repopulation by wildtype mosquitoes. Figure \ref{fig:ext_times} shows the conditional expected times until extinction of Wolbachia-only household populations with up to $30$ females calculated using a similar method to that described in Section \ref{subsec:absorbtime}; see Appendix \ref{ap:extinction_time} for more details. Expected extinction time increases with population size, but saturates. For population sizes at the deterministic steady state of $7$ Wolbachia-infected female mosquitoes, the extinction time is $167$ days, suggesting regular release may be needed to provide long term household protection. 

In the next section we will consider imperfect vertical transmission, so the offspring of Wolbachia-infected females are wildtype with probability $1-v$, where $v < 1$. This means a household in a state in the Wolbachia-only class can move to a state in the mixed class and subsequently to the wildtype-only class. We will call this reversion. Note that reversion is a separate process to the complete extinction of Wolbachia-infected mosquitoes and subsequent reconolisation by a new population of wildtypes. However, the random walk of the Wolbachia-infected population towards extinction can work synergistically with imperfect vertical transmission to speed up reversion.

\section{Reversion to wildtype}\label{sec:reversion}
\subsection{Model setup}\label{subsec:rev_setup}
If vertical transmission is imperfect ($v<1$) then some offspring of Wolbachia-infected mosquitoes are wildtype and states in $\mcs{3}$ are accessible from $\mcs{2}$, and vice-versa. So the state space $\mathcal{S}$ decomposes into two communicating classes. We relabel the communicating classes such that $\mcs{1}=\{(m,0):0 < m\leq 30\}$ contains the wildtype-only states and $\mcs{2}=\{(m,w): w>0,\; 0<m+w\leq 30\}$ contains the Wolbachia-only and mixed states. The full $\bm{Q}$ matrix in lower block triangular form is now
\begin{align}
    \bm{Q} = \begin{pmatrix}
        \bm{Q}_1 & \bm{0} \\
        \bm{Q}_{21} & \bm{Q}_2 \\
    \end{pmatrix},
\end{align}
where $\bm{Q}_1$ and $\bm{Q}_2$ are the sub-matrices representing the communicating classes $\mcs{1}$ and $\mcs{2}$ and $\bm{Q}_{21}$ contains the rates of transition from states in $\mcs{2}$ to states in $\mcs{1}$. The household is now able to escape the Wolbachia-only states into the mixed states, or the absorbing state, but remains unable to leave the wildtype-only class except to the absorbing state. We are interested in the probability that a household that has reached the Wolbachia-only class returns to the mixed class. We call this the reversion probability.

\subsection{Results}\label{subsec:6_2}
\begin{figure}
    \centering
    \includegraphics[width=0.5\linewidth]{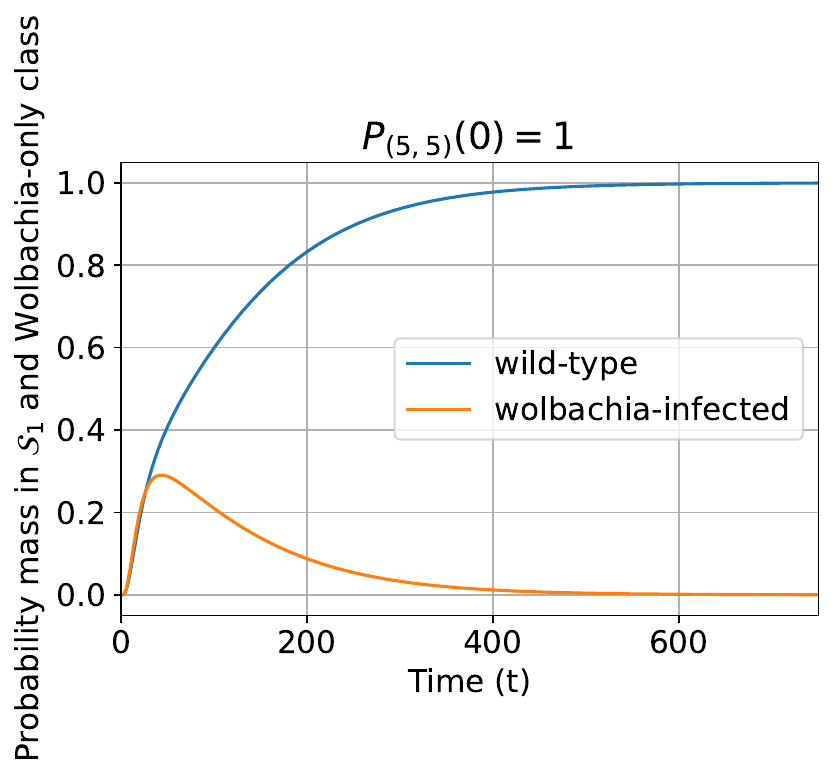}
    \caption{Dynamics of the probability distribution for the wildtype-only and Wolbachia-only classes conditioned on non-extinction when vertical transmission is imperfect. Note that the Wolbachia-only class is not a communicating class here. The class probability is the sum of the probabilities for all states in that class. For the model with households of up to $30$ female mosquitoes, $v=0.9$, $1-\phi=0.15$, larval density function is \eqref{eq:hughes_F}. Initially $P_{(5,5)}(0)=1$.}
    \label{fig:prob_cond_Dye_rev}
\end{figure}

We set the vertical transmission probability to $v=0.9$ but keep all other parameters as in Section \ref{sec:30mosq}. The minimal magnitude eigenvalues of the two communicating classes are given in Table \ref{tab:eigs}. The overall minimal magnitude eigenvalue is unchanged, and associated with class $\mcs{1}$, for any value of $v$. So because the dynamics of the chain within class $\mcs{1}$ do not depend on $v$, the QSD is the same as for the model without reversion. However, the QSD is reached much more quickly than when $v=1$, see Figure \ref{fig:prob_cond_Dye_rev}. This is because the leaky vertical transmission increases the effective reproduction rate for wildtype mosquitoes, and reduces it for Wolbachia-infected mosquitoes. Consequently, Wolbachia-only states move towards extinction more quickly. 

The damping ratios for this model are given in Table \ref{tab:trans_results}. The state entropies are shown in Figure \ref{fig:entropies_hughes_Dye_v} in Appendix \ref{ap:reversion} alongside results for the probability of Wolbachia invasion and the expected time to invasion. These quantities are very similar to the $v=1$ case.

\subsubsection{Reversion probability} \label{subsec:rev_prob}
The reversion probability and expected time until reversion takes place can be used to estimate the duration of household protection and inform the release frequency of Wolbachia-infected mosquitoes. The probability of reversion can be found using the methodology described in Section \ref{subsec:absorbprob}. We now have two communicating classes and need to find a column vector $\bm{a}(m',0)=\left(a_{(m,w)}(m',0):(m,w)\in \mcs{2}\right)$ containing the probabilities of entering the wildtype-only class at state $(m',0)$ from each state in $\mcs{2}$ by solving the linear system
\begin{align}\label{eq:rev_prob}
    -\bm{Q}_2\bm{a}(m',0)=\bm{q}(m',0),
\end{align}
where $\bm{Q}_2$ is the sub-Q matrix for $\mcs{2}$ and $\bm{q}(m',0)$ is a column vector of the transition rates from each state in $\mcs{2}$ to $(m',0)$. The probability of reaching $\mcs{1}$ from a given state $(m,w)$ in $\mcs{2}$ is the sum of $a_{(m,w)}(m',0)$ over all $(m',0) \in \mcs{1}$. We determine the total probability of reaching $\mcs{1}$ from a given state distribution in $\mcs{2}$ by using that distribution to construct a weighted sum of the probabilities of reaching $\mcs{1}$ from each $(m,w) \in \mcs{2}$. 

\begin{figure}
    \centering
    \includegraphics[width=0.5\linewidth]{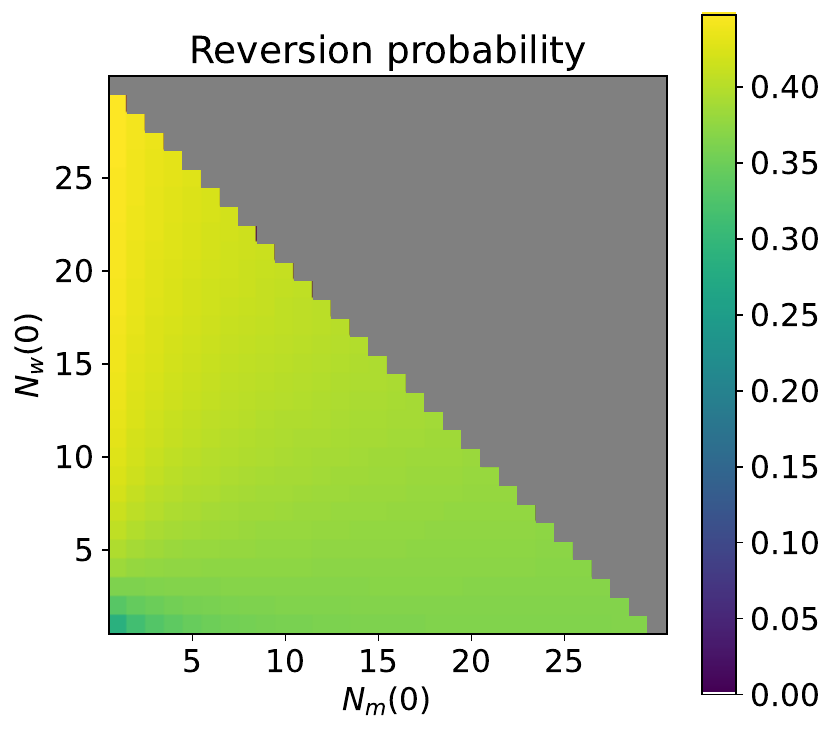}
    \caption{Probability of reversion from the Wolbachia-only states to the mixed states when a Wolbachia-only state is reached from an initial mixed state $(N_m(0), N_w(0))$. For the model with households of up to $30$ mosquitoes, $v=0.9$, $1-\phi=0.15$, larval density function is \eqref{eq:hughes_F}.}
    \label{fig:prob_reinvade-dye}
\end{figure}

To determine the state distribution of $\mcs{2}$, consider that the wildtype-only class may be re-entered via any state in $\mcs{2}$ and the Wolbachia-only states in $\mcs{2}$ will have first passed through some mixed state $(m,w)$. Therefore if we condition on starting at mixed state $(m,w)$ with probability $1$, it is possible to obtain the state distribution in the Wolbachia-only state space by considering the probabilities of entering each Wolbachia-only state (first out of all the Wolbachia-only states) via the mixed state $(m,w)$. These probabilities are found by solving equations \eqref{eq:rev_prob} with $(m',0)$ swapped for Wolbachia-only state $(0,w')$ and only taking the rows and columns of $\bm{Q}_2$ corresponding to the mixed states. This allows us to perform an informal sensitivity analysis on the probability of reversion occurring within the household, conditioning on starting in the mixed state $(m,w)$ via which the Wolbachia-only state space is first entered. This is presented in Figure \ref{fig:prob_reinvade-dye}.

Giving equal weight to each possible initial mixed state, the average reversion probability when $1-\phi=0.15$ was $0.39$. In the specific case where the Wolbachia-only class is reached from an initial mixed state $(5,5)$ the reversion probability is $0.38$. In general, the reversion probability is slightly lower when starting from mixed states with smaller populations of mosquitoes or higher proportions of wildtype mosquitoes. In these cases, there is a higher probabilty that the Wolbachia-only class is entered via a state with low numbers of Wolbachia-infected mosquitoes and thus there is a higher risk of extinction before reversion is possible.
  
\subsubsection{Expected time until reversion} \label{subsec:rev_time}
\begin{figure}
    \centering
    \includegraphics[width=0.5\linewidth]{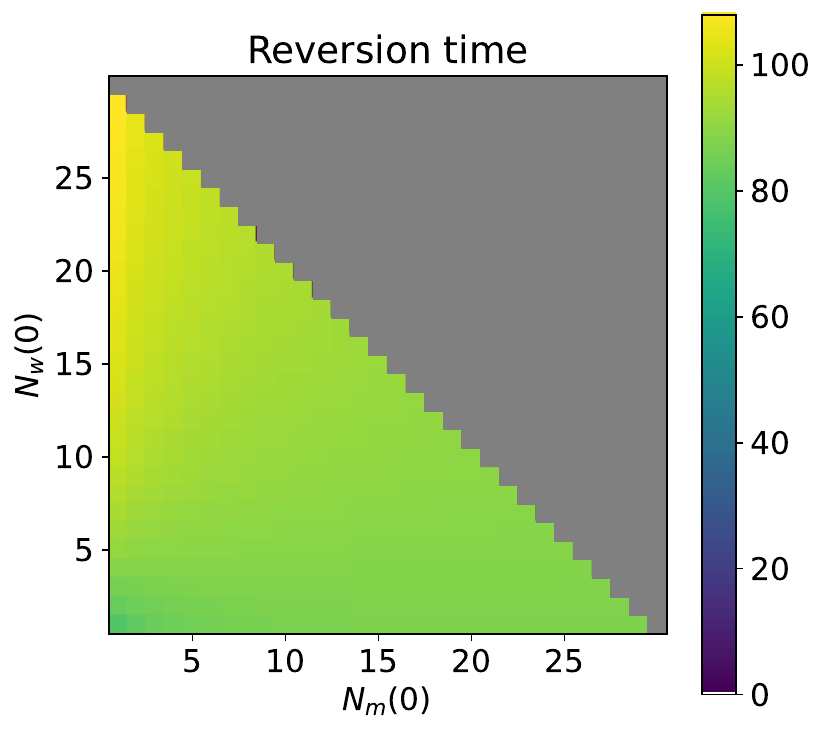}
    \caption{Expected time until reversion (in days) from the Wolbachia-only states to the mixed states (conditional on the event that reversion occurs) when a Wolbachia-only state is reached from an initial mixed state $(N_m(0), N_w(0))$. For the model with households of up to $30$ mosquitoes, $v=0.9$ $1-\phi=0.15$, larval density function is \eqref{eq:hughes_F}.}
    \label{fig:time_reinvade-dye}
\end{figure}

We perform the same sensitivity analysis on the expected time until reversion (conditional on the event that reversion occurs) from a Wolbachia-only state to a wildtype-only state (Figure \ref{fig:time_reinvade-dye}), conditioning on each possible mixed state $(m,w)$ initial condition. The average expected reversion time is $92$ days and the reversion time conditional on the mixed state initial condition $(5,5)$ is $89$ days. When the initial mixed state has a high proportion of Wolbachia-infected mosquitoes, the expected time until reversion after entering the Wolbachia-only class is slightly higher. This is because there is a higher probability that the Wolbachia-only class is entered via a state with high numbers of Wolbachia-infected mosquitoes. A large resident Wolbachia-infected population reduces the risk of extinction and, via density dependent competition and CI-induced inhibition of wildtype reproduction, the risk of re-invasion due to imperfect vertical transmission. 

\section{Discussion}
In this study, we developed a stochastic model for the population dynamics of wildtype and Wolbachia-infected mosquitoes in a household setting. We used a continuous-time Markov chain framework aligned closely with a previously published mean-field model \cite{hughes2013modelling}. Our overarching aim was to understand how deterministic bistability between wildtype-only and Wolbachia-only steady states, which determines whether a Wolbachia release leads to a successful Wolbachia invasion, is impacted by the inherent stochasticity of small populations.

Starting with a simplified tutorial model, we were able to utilise and adapt Markov and matrix population model theory to gain insights into the transient and asymptotic dynamics of the stochastic system. 
We calculated quantities such as the probability of Wolbachia invasion into a household after a release has been carried out, the expected time until invasion, and the quasi-stationary distribution of the system. This information provides estimates of the number of Wolbachia-infected mosquitoes that must be released in order to displace the wildtype population, and the frequency with which releases must be repeated in order to sustain the Wolbachia infection within the household. We focused on the impact of two key parameters: the fitness cost of Wolbachia infection on the mosquito's birth rate, $1-\phi$, and the probability of vertical transmission of Wolbachia infection, $v$, which are known to vary between Wolbachia strains and have been shown in deterministic models to be important factors affecting the establishment of Wolbachia infection. 

We studied the Wolbachia invasion probability under a range of initial population sizes of wildtype and Wolbachia-infected mosquitoes. We observed that the invasion probability is relatively consistent along the deterministic invasion boundary, but is quite small. 

We compared frameworks that omit and allow reversion to a wildtype-only population after the Wolbachia-infected population has become established. This occurs via Wolbachia-infected females producing some wildtype offspring due to imperfect vertical transmission. Literature suggests the vertical transmission probability $v$ is close to $1$. We observed a difference in communicating class structure between the two frameworks. In reality, wildtype populations may be able to re-invade in either case if they migrate in from elsewhere, such as surrounding rural areas. 

We found that introducing the minimum number of Wolbachia-infected mosquitoes required to entirely displace the wildtype population (ranging between $0$ and $25$ wildtypes) in the deterministic model only achieves that outcome approximately $23$\% of the time in the stochastic model ($v=1$). Successfully displacing the wildtype population at least $90$\% of the time requires the introduction of a much larger number of Wolbachia-infected mosquitoes. In the deterministic model we observe a hard boundary between the basins of attraction of the wildtype-only and Wolbachia-infected only steady states. But that boundary is blurred in the stochastic model because small populations of Wolbachia-infected mosquitoes often become extinct before proliferating sufficiently for cytoplasmic incompatibility to suppress the wildtype population. 

Since Ae. aegypti mosquitoes often live in and around the dwellings of the people that they bite, and may also be released at a household scale, their invasion dynamics will unfold stochastically within small populations. Therefore, it is imperative to assess invasion probabilities using stochastic methods that account for spatial structure at a household scale.

We acknowledge several limitations and assumptions in our model. We only explicitly modelled the female mosquito population, and assumed there are an equal number of males in the population. This approximation is in line with deterministic models \cite{hughes2013modelling,qu2022modeling} and is reasonable because we examine only the simple setting of a single household, with the objective of realising the mosquito invasion dynamics at a household level. However, male only releases are a common control strategy for mosquito population suppression, acting similarly to the sterile insect technique \cite{ross2021designing}. We imposed an upper bound on the female population size in the household of $30$ mosquitoes. This assumption is necessary because otherwise the state space $\mathcal{S}$ would become infinite. Although we suspect many of our analytical quantities would still hold under an infinite state space $\mcs{}$, the numerical results from this study would become unattainable. Observed household mosquito populations reported in the literature suggest this upper bound is reasonable. We only considered the invasion dynamics in a single household or, equivalently, a community composed of a large number of independent households. This assumption allows analytic tractability and ensures the state space is computationally manageable. However, movement of mosquitoes between households may be an important factor in the establishment and maintenance of Wolbachia infection at the household and community scales. 

In future work we will address some of these issues. We will extend the single household framework to a system of connected households. This will require the incorporation of movement of mosquitoes between households. Male and female Ae. aegypti movement behaviour is contrasting. Males tend to move around more frequently in search of mates \cite{trewin2021mark}, moving distances of around $200$ to $400$ metres over their lifetime \cite{juarez2020dispersal,marcantonio2019quantifying,trewin2020urban}, whereas females typically move less than $100$ metres in their lifetime. This distinction reinforces the need to consider differentiating between male and female mosquitoes. It is also possible that sex ratios among wildtype and infected mosquito types may become skewed, altering the reproductive rates and driving qualitative changes in the invasion dynamics. A community of connected households may also affect the bistability result reported for a single population if there is a larger set of possible steady state solutions. Incorporating movement between households into our CTMC model will introduce nonlinearities related to community scale dispersal to household scale containers. This complexity will stymie many of the analytic approaches discussed in this study, and necessitate the use of stochastic simulation methods \cite{gillespie1977exact,barlow2024epidemiological}.

We defined a larval density dependence function acting on the birth rate, which describes the effect of intraspecific competition for resources such as food and space. We investigated two common larval density dependence functions \cite{dye1984models,qu2022modeling}. Both of these models are heuristic, although some data was used to estimate parameters in equation \eqref{eq:hughes_F} under the original scaling \cite{hughes2013modelling,dye1984models}. The overall results of our analysis are qualitatively similar for both models. The main difference is that equation \eqref{eq:hughes_F} produces a much higher birth rate than \eqref{eq:qu}, leading to the Wolbachia-only steady state existing (in the mean-field model) for a larger region of the Wolbachia fitness parameter space. Were more data to become available, it would beneficial to revisit the larval density dependence function used here and in other studies \cite{dye1984models,hughes2013modelling,ndii2012modelling,qu2022modeling}. 

In this study we have concentrated on the mosquito invasion dynamics alone. But the overarching motivation for establishing Wolbachia infection in domestic mosquitoes is to reduce the circulation of dengue and other vector-borne diseases. In many cases the host-vector transmission dynamics of those infectious diseases will also take place at a household scale and differences in human household sizes and commuter patterns could act to transform the mosquito invasion dynamics. There has been some progress in modelling vector-borne disease dynamics at the household scale \cite{black2022stochastic}. An obvious next step would be to develop that framework to include Wolbachia infection. 

In conclusion, Aedes aegypti mosquitoes favour urban residential habitats. The mosquito population dynamics, and efforts to control them, unfold stochastically at a household scale. Deterministic models have demonstrated that releasing Wolbachia-infected mosquitoes can displace the wildtype population, as long as a sufficient number of infected mosquitoes are introduced. Stochastic models support this inference, but suggest that the relationship between the initial introduction and the eventual outcome is more nuanced.   

\backmatter
\bmhead{Supplementary information}
All code used to produce the results presented in this paper can be found at \url{https://github.com/ahb48/Wolbachia_invasion_households}. %Supplementary information can be found in a supporting document online.

\bmhead{Acknowledgements}
This work was supported by a scholarship from the EPSRC Centre for Doctoral Training in Statistical Applied Mathematics at Bath (SAMBa), under the project EP/S022945/1. We thank Alex Cox for his advice on formalising Appendix sections \ref{ap:claim1} and \ref{ap:claim2}.

For the purpose of open access, the author has applied a Creative Commons Attribution (CC-BY) license to any Author Accepted Manuscript version arising.
No new data was created during the study.

\begin{appendices}
\section{Alternative larval density dependence function}\label{ap:dye}
All results in the main text come from models that use the function $F(N)$ for density dependent competition at the larval stage where
\begin{align}
    F(N)=\begin{cases}\exp(-hN^k) & N < C,\\
    0 & N \geq C.
    \end{cases}
\end{align}
This function was first introduced in \cite{dye1984models}, and used in many subsequent studies including \cite{hughes2013modelling}. A simpler alternative to $F(N)$ is $G(N)$ given by \cite{qu2022modeling} 
\begin{align}\label{eq:qu}
    G(N) = \begin{cases} 
    1-N/C & N < C,\\
    0 & N \geq C.
    \end{cases}
\end{align}
Here we reproduce our main results using $G(N)$. We use the same parameter regime as in Table \ref{tab:params}. As before, we determine the per capita birth rate $b$ by fixing the wildtype-only steady state at $x^*=10$ and solving $bx^*G(x^*)=dx^*$. The shape of $G(N)$ results in $b=0.18$, much lower than the $b=0.54$ obtained with $F(N)$.

\subsection{Bistability in the mean-field model}
\begin{figure}
    \centering
    \includegraphics[width=0.5\linewidth]{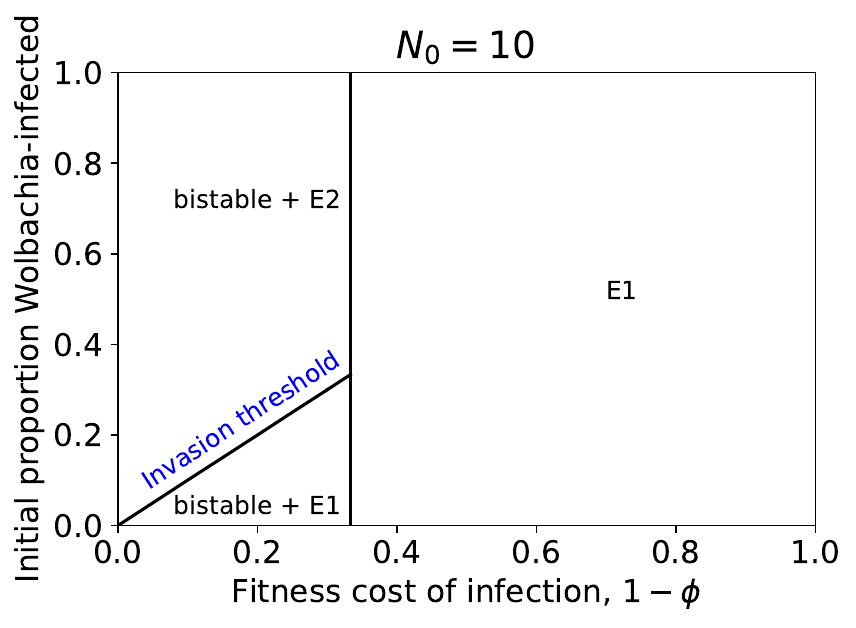}
    \caption{Stability and attraction of the steady states of system \eqref{eq:hughes}--\eqref{eq:hughes2}, with larval density function \eqref{eq:qu}, depending on the fitness cost of Wolbachia infection $1-\phi$ and the initial proportion of the mosquito population that is Wolbachia-infected $N_w(0)/N_0$. Here $u,v=1$ and $N_0 = 10$. Indicated on the diagram are the regions where the system is bistable and either the wildtype-only ($E1$) or Wolbachia-only ($E2$) steady state is attracting, and where the unique non-trivial steady state is wildtype-only, and stable. The invasion threshold under the bistable region is labelled in blue text. All of our stability analysis plots are produced using the open source software bSTAB \cite{stender2022bstab}.}\label{fig:basin_attr}
\end{figure}

Figure \ref{fig:basin_attr} shows the stability and basins of attraction of the steady states of the mean-field model \eqref{eq:hughes}--\eqref{eq:hughes2}. As before (Figure \ref{fig:basin_attr_Dye}) we observe a bistable region. However, the lower birth rate associated with $G(N)$ means that the bistable region is much smaller as the Wolbachia-only steady state only exists when $1-\phi<0.33$. 

\begin{figure}
    \centering
    \includegraphics[width=0.5\linewidth]{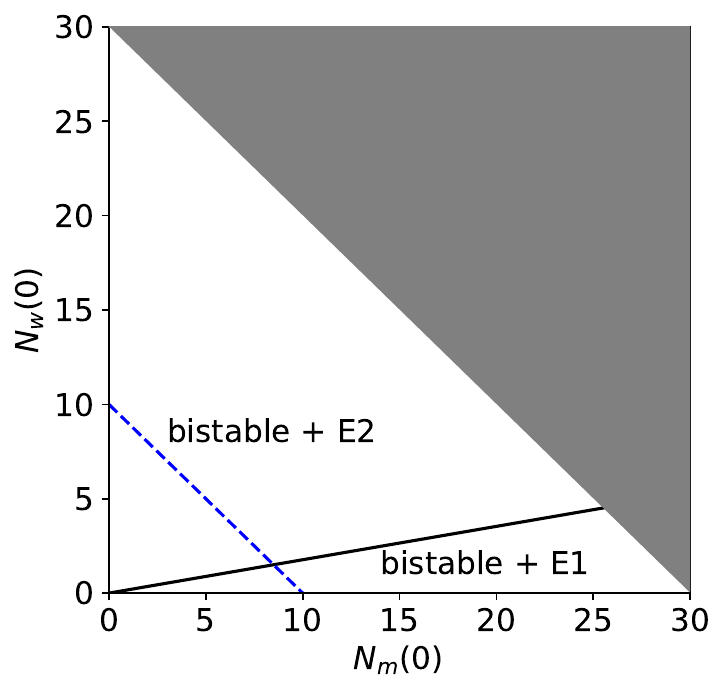}
  \caption{Stability and attraction of steady states of system \eqref{eq:hughes}--\eqref{eq:hughes2} depending on the initial numbers of female wildtype $N_m(0)$ and Wolbachia-infected $N_w(0)$ mosquitoes. Here $u,v=1$ and $1-\phi=0.15$, with larval density function \eqref{eq:qu}. Indicated on the diagram are the regions where the wildtype-only ($E1$) or Wolbachia-only ($E2$) steady states are attracting. The grey triangle covers populations outside our upper bound on the total female mosquito population size. The dashed blue line indicates where $N_m(0)+N_w(0)=10$, corresponding to $1-\phi = 0.15$ in Figure \ref{fig:basin_attr}.}
  \label{fig:stab_colour}
\end{figure}

Figure \ref{fig:stab_colour} shows a basin of attraction across the full set of feasible initial conditions at a parameter location in the bistable region. The alternative density dependence function $G(N)$ does not significantly change the results (compare with Figure \ref{fig:stab_colour_Dye}). 

\subsection{State distributions and invasion probabilities in the stochastic model, $v=1$}\label{subsec:norev_Dye}
\begin{figure}
    \centering
    \includegraphics[width=0.5\linewidth]{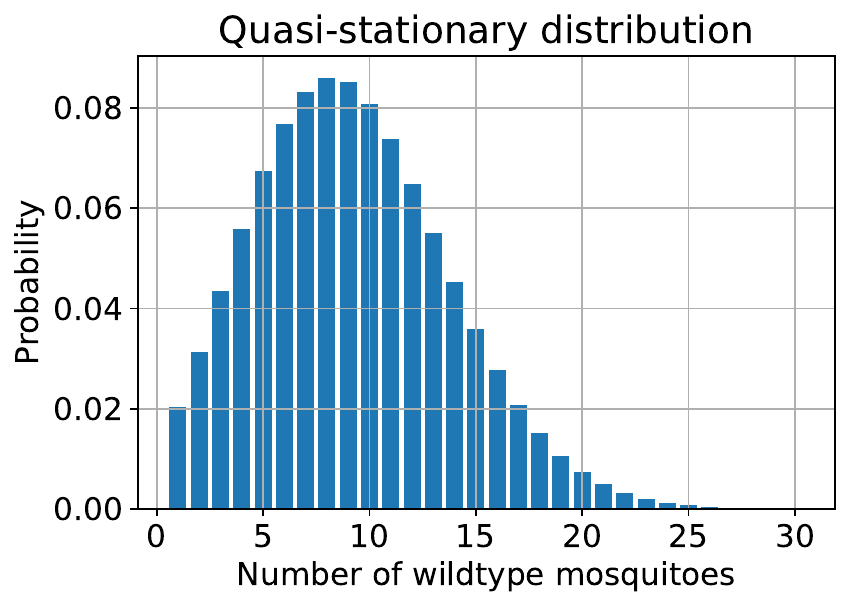}
    \caption{QSD for the model with households of up to $30$ female mosquitoes, perfect vertical transmission ($v=1$), $1-\phi=0.15$ and larval density function \eqref{eq:qu}. Any states not shown have probability equal to $0$.}
    \label{fig:qsd30}
\end{figure}

The larval density dependence function does not affect the structure of the communicating classes (Section \ref{sec:30mosq}). The overall minimal magnitude eigenvalue, and hence the QSD, is still associated with the wildtype-only class. Figure \ref{fig:qsd30} shows the QSD (which has all of its mass in the wildtype-only class). The alternative density dependence function $G(N)$ has little impact on the QSD (compare with Figure \ref{fig:qsd_Dye}) except that the tail to the right is slightly smaller due to the lower birth rate.

\begin{figure}
    \centering
    \includegraphics[width=0.5\linewidth]{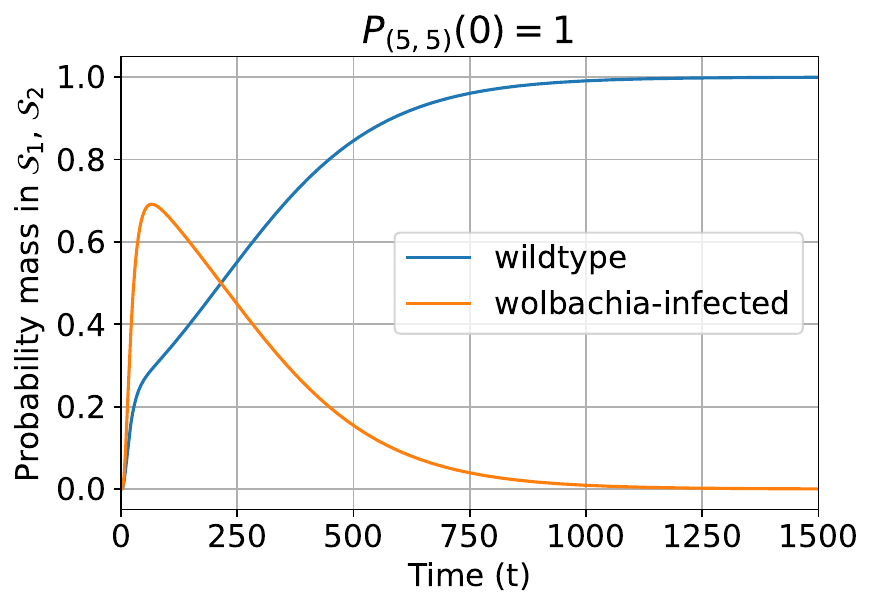}
    \caption{Dynamics of the probability distribution for the wildtype-only and Wolbachia-only classes conditioned on non-extinction. The class probability is the sum of the probabilities for all states in that class. For the model with households of up to $30$ mosquitoes, perfect vertical transmission ($v=1$), $1-\phi=0.15$ and larval density function \eqref{eq:qu}. Initially $P_{(5,5)}(0)=1$. For small $t$ some of the probability mass is in the mixed class $\mcs{3}$ but this quickly decays and has been omitted here.}
    \label{fig:prob_dist_hughes_comp}
\end{figure}

Figure \ref{fig:prob_dist_hughes_comp} shows the probabilities over time of a household being in the wildtype-only and Wolbachia-only classes, conditioned on non-extinction and initial state probability distribution $P_{(5,5)}(0)=1$. The alternative density dependence function $G(N)$ has little impact on the distribution (compare with Figure \ref{fig:prob_cond_Dye}) but the distribution converges slightly faster to the QSD. This observation is corroborated by the damping ratio.

\begin{figure}
    \centering
    \includegraphics[width=0.7\linewidth]{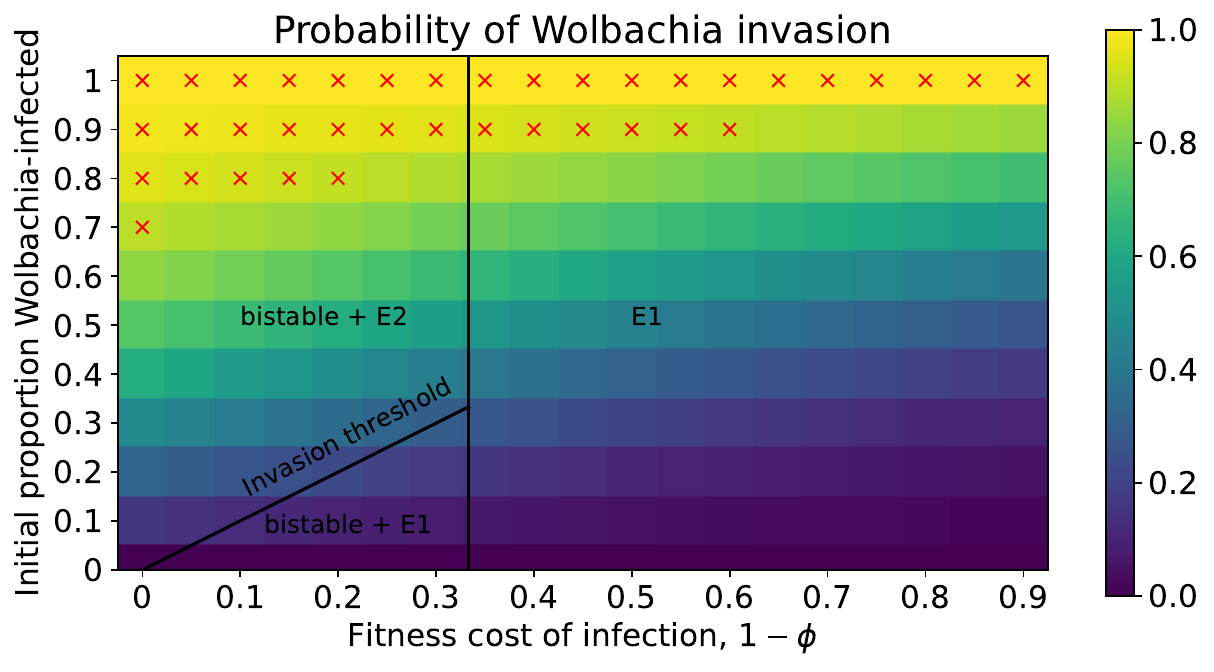}
    \caption{Probability of successful Wolbachia invasion of a household depending on the initial Wolbachia-infected proportion $N_w/N_0$ and fitness cost of infection $1-\phi$ with larval density function \eqref{eq:qu}. The initial household size is fixed at $N_0=10$ female mosquitoes and the initial infected proportion is determined by $N_w(0)$ with $N_m(0) + N_w(0) = N_0$. Households can have up to $30$ female mosquitoes and $v=1$. The solid lines and text show the invasion threshold and stable steady states of the corresponding deterministic model. The red crosses indicate where the probability of successful Wolbachia invasion is greater than or equal to $0.9$.}
\label{fig:color_plot_invasion_prob_hughes_comp}
\end{figure}

\begin{figure}
  \centering
  \begin{subfigure}{0.4\textwidth}
    \includegraphics[width=\linewidth]{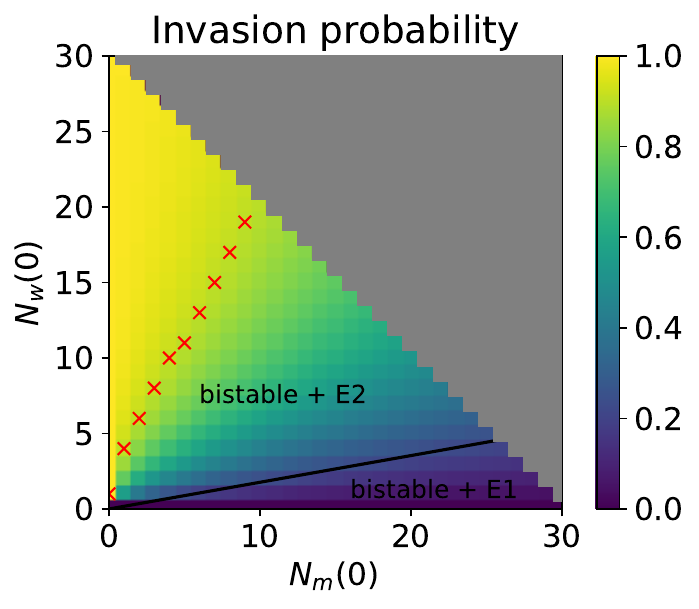}
    \caption{}
    \label{fig:subfig1_prob_qu_v1}
  \end{subfigure}
  \hfill
  \begin{subfigure}{0.4\textwidth}
    \includegraphics[width=\linewidth]{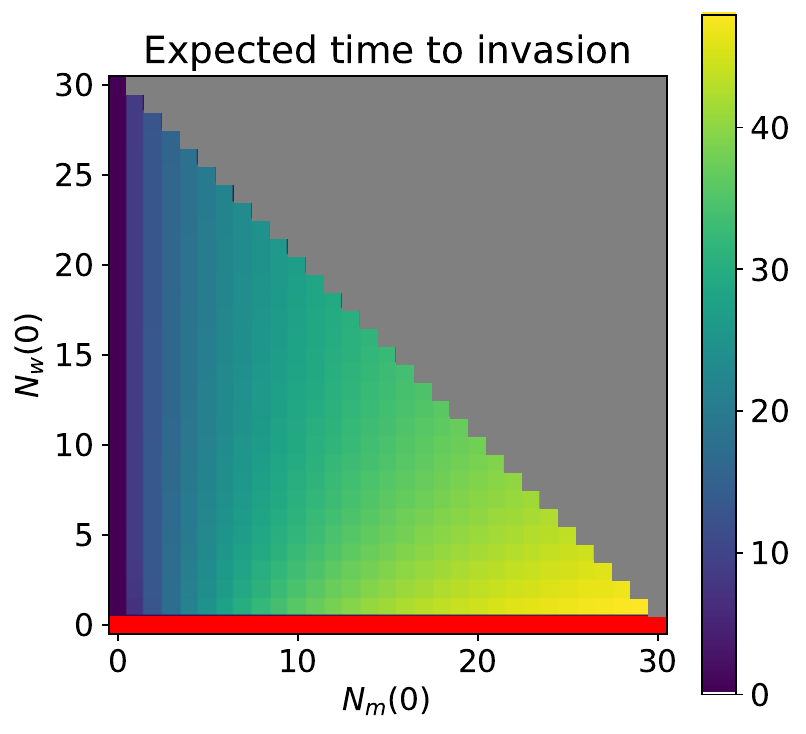}
    \caption{}
    \label{fig:subfig2_absorb_time_qu_v1}
  \end{subfigure}
  \caption{(a) Probability of entering the Wolbachia-infected class when the initial state is $(N_m(0), N_w(0))$. The solid black line is the corresponding deterministic threshold (Figure \ref{fig:stab_colour}). Red crosses denote, for different wildtype mosquito numbers, the minimum number of Wolbachia-infected mosquitoes required for successful invasion with probability $0.9$ or higher. Note that, if there are more than $9$ wildtype female mosquitoes present, the number of Wolbachia-infected mosquitoes required for a successful invasion with probability $0.9$ or higher is outside of the state space. (b) Expected time in days to enter the Wolbachia-infected class, conditional on this event occurring, when the initial state is $(N_m(0), N_w(0))$. The red bar indicates initial conditions where only wildtype mosquitoes are present and so the Wolbachia-infected class cannot be reached. In both diagrams the maximum number of female mosquitoes in a household is $30$, the initial state probability distribution is $P_{(N_m(0),N_w(0))}(0)=1$, parameters are as in Table \ref{tab:params}, $1-\phi=0.15$, $v=1$, and the larval density function is \eqref{eq:qu}. The top right quadrant shaded grey is not an admissible region of the state space.}\label{fig:color_plots_hughes_comp}
  \end{figure}

Figure \ref{fig:color_plot_invasion_prob_hughes_comp} shows how the fitness cost of infection $1-\phi$  affects the probability of Wolbachia invasion (i.e. reaching a Wolbachia-only state) in a household containing a total of $N_0=10$ female mosquitoes, of which a proportion $N_w(0)/N_0$ are Wolbachia-infected. The alternative density dependence function $G(N)$ has little impact on the correspondence between the deterministic bistability and the stochastic invasion probability (compare with Figure \ref{fig:color_stab_dye_v1}). As such, $G(N)$ tends to shift regions with high probability of reaching the Wolbachia-only state to lower values of $1-\phi$. There is little change between Figures \ref{fig:subfig1_prob_invade} and \ref{fig:subfig1_prob_qu_v1}. However, the expected time until Wolbachia invasion increases overall (compare Figures \ref{fig:subfig2_time_invade} and \ref{fig:subfig2_absorb_time_qu_v1}) due to the decrease in birth rate.

\begin{figure}
    \centering
    \includegraphics[width=0.5\linewidth]{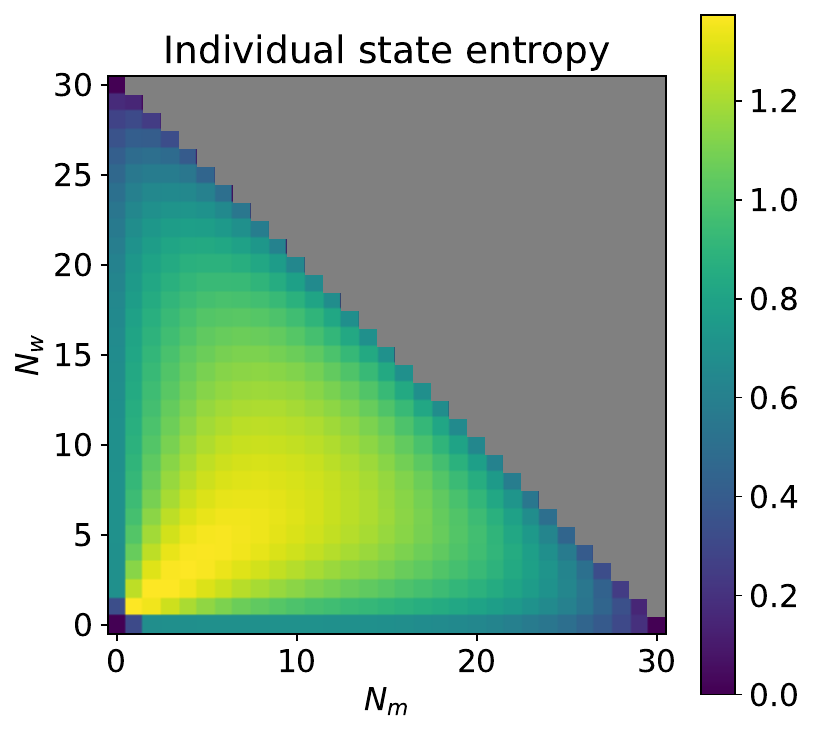}
    \caption{Entropy of state $(N_m,N_w)$. Higher entropy corresponds to greater uncertainty in the next state. The maximum number of female mosquitoes in household is $30$, parameters are as in Table \ref{tab:params}, $1-\phi=0.15$, $v=1$, larval density function is \eqref{eq:qu}. The top right quadrant shaded grey is not an admissible region of the state space.}
    \label{fig:entropies_hughes_comp}
\end{figure}

Figure \ref{fig:entropies_hughes_comp} shows the individual state entropies, a measure of the uncertainty in the next state to be visited. The alternative density dependence function $G(N)$ has little impact on the state entropies, except close to the boundary
$N_m+N_w = C$, where they are markedly reduced (compare with Figure \ref{fig:entropies_hughes_Dye}). This reduction is due to a much stronger larval density effect close to the maximum population size $C$ (see Figure \ref{fig:larval_density_funcs}) which reduces the probability of a birth occurring in states close to the boundary. 

\begin{figure}
    \centering
    \includegraphics[width=0.5\linewidth]{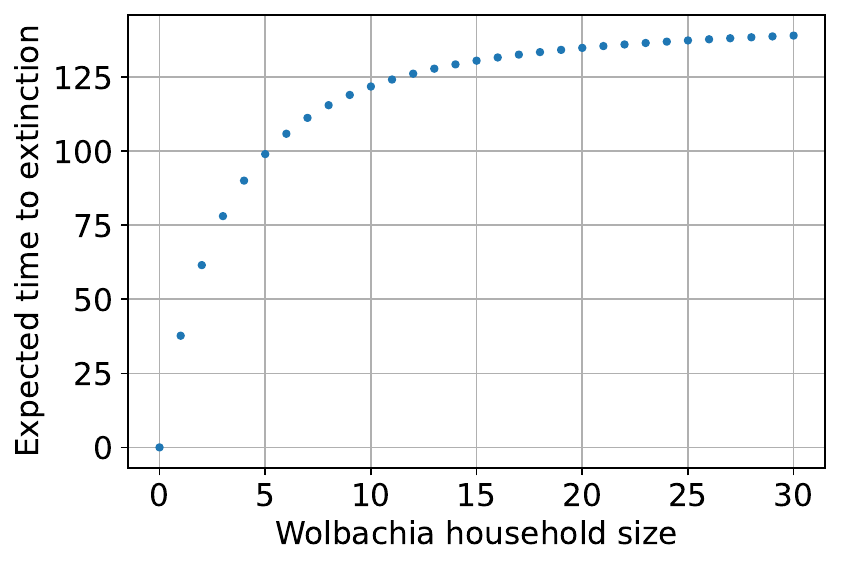}
    \caption{Expected time until extinction for a household population with no wildtype mosquitoes and up to $30$ Wolbachia-infected female mosquitoes. Parameters are as in Table \ref{tab:params}. The fitness cost of Wolbachia infection is $1-\phi=0.15$, $v=1$ and larval density function is \eqref{eq:qu}.}
    \label{fig:ext_times_qu}
\end{figure}

Figure \ref{fig:ext_times_qu} shows the expected time until extinction for a Wolbachia-only household. The alternative density dependence function $G(N)$ reduces the time to extinction (compare with Figure \ref{fig:ext_times}). This reduction is a consequence of the lower birth rate.

\subsection{State distributions and invasion probabilities in the stochastic model with reversion, $v=0.9$}
We next present results for the model with households of up to $30$ female mosquitoes with parameters comparable to the mean-field model \eqref{eq:hughes}--\eqref{eq:hughes2} where vertical transmission is imperfect, $v=0.9$. As above, we use the larval density dependence function \eqref{eq:qu} and set the fitness cost of Wolbachia infection to $1-\phi=0.15$ unless otherwise stated. The communicating class structure is the same as in Section \ref{sec:reversion}.
The QSD is once again identical to the QSD in the previous Section \ref{subsec:norev_Dye} (see Section \ref{subsec:6_2}), so is omitted.

\begin{figure}
    \centering
    \includegraphics[width=0.5\linewidth]{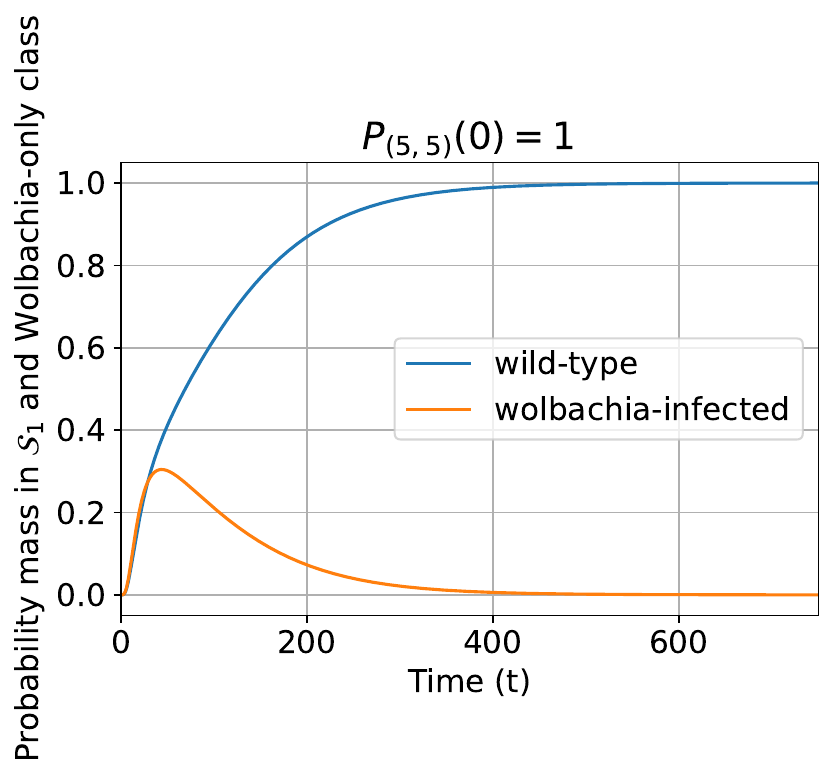}
    \caption{Dynamics of the probability distribution for the wildtype-only class  and Wolbachia-only classes conditioned on non-extinction when vertical transmission is imperfect.  Note that the Wolbachia-only class is not communicating here. The class probability is the sum of the probabilities for all states in that class. For the model with households of up to $30$ female mosquitoes, $v=0.9$, $1-\phi=0.15$, larval density function is \eqref{eq:qu}. Initially $P_{(5,5)}(0)=1$.}
    \label{fig:Hughes_comp_prob_cond}
\end{figure}

The probability distribution of the household states over time (conditioned on non-extinction and $P_0(5,5)=1$; Figure \ref{fig:Hughes_comp_prob_cond}) is similar to that of Figure \ref{fig:prob_cond_Dye_rev} but converges slightly faster to the QSD. %This is corroborated by the damping ratio.

\begin{figure}
    \centering
    \includegraphics[width=0.5\linewidth]{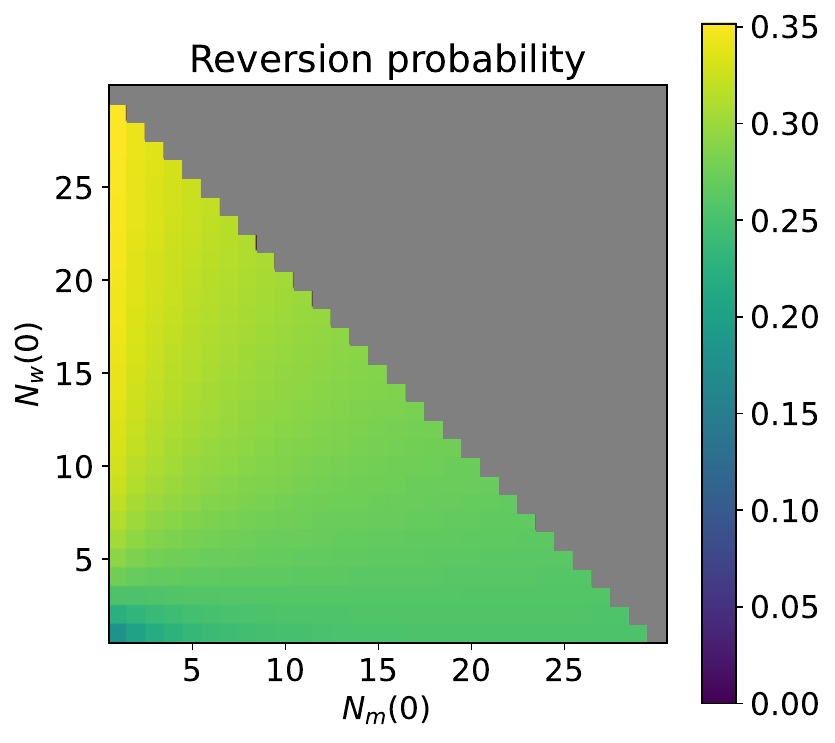}
    \caption{Probability of reversion from the Wolbachia-only class to the mixed class when the Wolbachia-only class is reached from an initial mixed state $(N_m(0), N_w(0))$. For the model with households of up to $30$ mosquitoes, $v=0.9$, $1-\phi=0.15$, larval density function is \eqref{eq:qu}.}
    \label{fig:prob-reinvasion-qu}
\end{figure}

\begin{figure}
    \centering
    \includegraphics[width=0.5\linewidth]{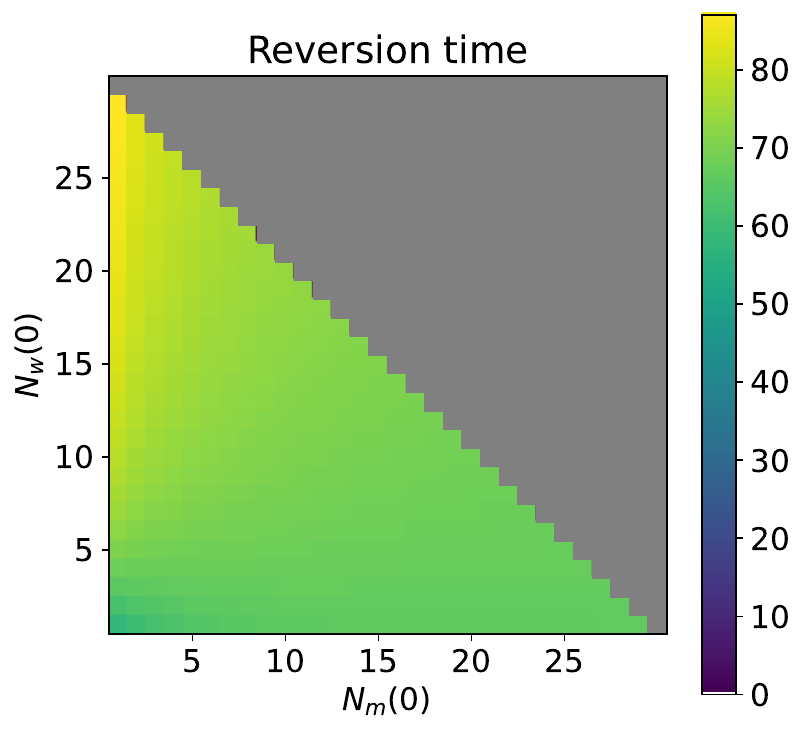}
    \caption{Expected time until reversion (in days) from the Wolbachia-only class to the mixed class (conditional on the event that reversion occurs) when the Wolbachia-only class is reached from an initial mixed state $(N_m(0), N_w(0))$. For the model with households of up to $30$ mosquitoes, $v=0.9$, $1-\phi=0.15$, larval density function is \eqref{eq:qu}.}
    \label{fig:time-reinvasion-qu}
\end{figure}

Repeating the sensitivity analysis we carry out in Section \ref{sec:reversion} with the alternative larval density dependence function $G(N)$, we obtain similar results (Figure \ref{fig:prob-reinvasion-qu} and Figure \ref{fig:time-reinvasion-qu}), although the reversion probability is slightly lower. The average reversion probability conditioned on initially reaching the Wolbachia-only state from an initial mixed state $(m,w)$ for $0<m,w\leq 30$ is $0.29$ and the reversion probability assuming we enter the Wolbachia-only class via $(5,5)$ is $0.27$. These lower values are due to the decrease in wildtype birth rate. The expected time until reversion (conditional on the event that reversion occurs) also decreases. The average conditional expected time until reversion is $71$ days and the conditional expected time to reversion from a critical mixed state $(5,5)$ is $68$ days. 
  
\section{Invasion probability with population size unconserved}\label{ap:pop_conserve}
\begin{figure}
    \centering
    \includegraphics[width=0.7\linewidth]{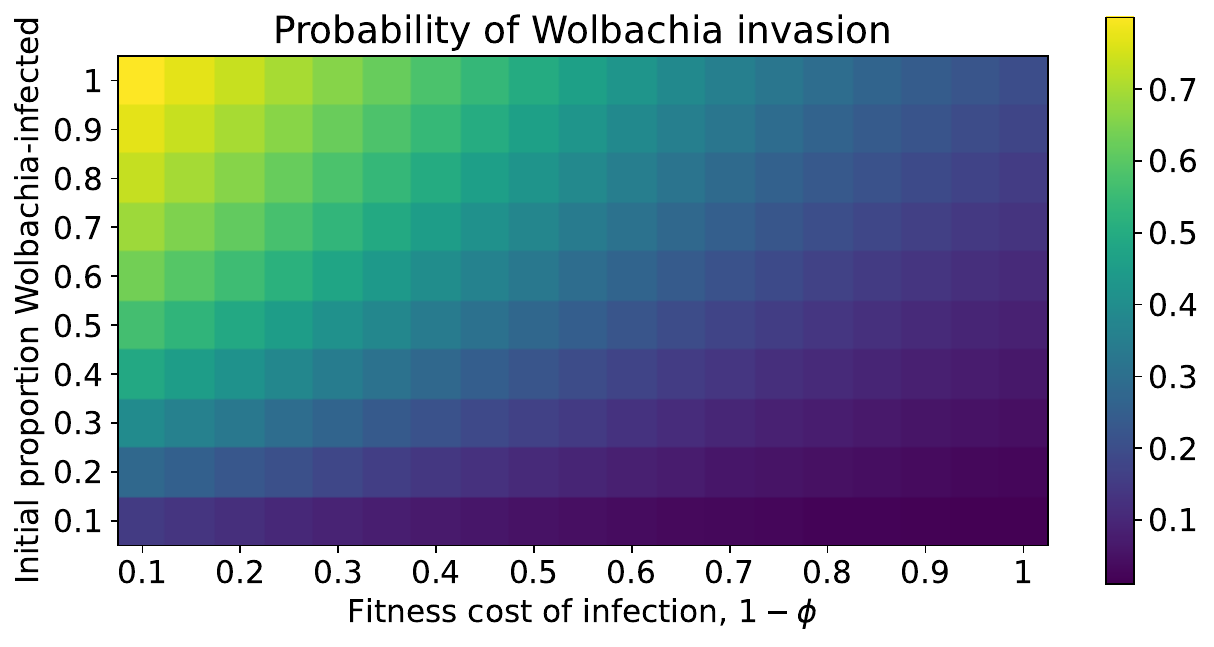}
    \caption{Probability of successful Wolbachia invasion of a household depending on the initial (female) Wolbachia-infected proportion $N_w/N_0$ and fitness cost of infection $1-\phi$. Initially the household is composed of $N_0=10$ female wildtype mosquitoes and between $1$ and $10$ additional Wolbachia-infected females are introduced so the total household population is between $11$ and $20$. Households can have up to $30$ female mosquitoes and $v=1$ and the larval density function is \eqref{eq:hughes_F}.}
    \label{fig:hughes_v1}
\end{figure}

\begin{figure}
    \centering
    \includegraphics[width=0.7\linewidth]{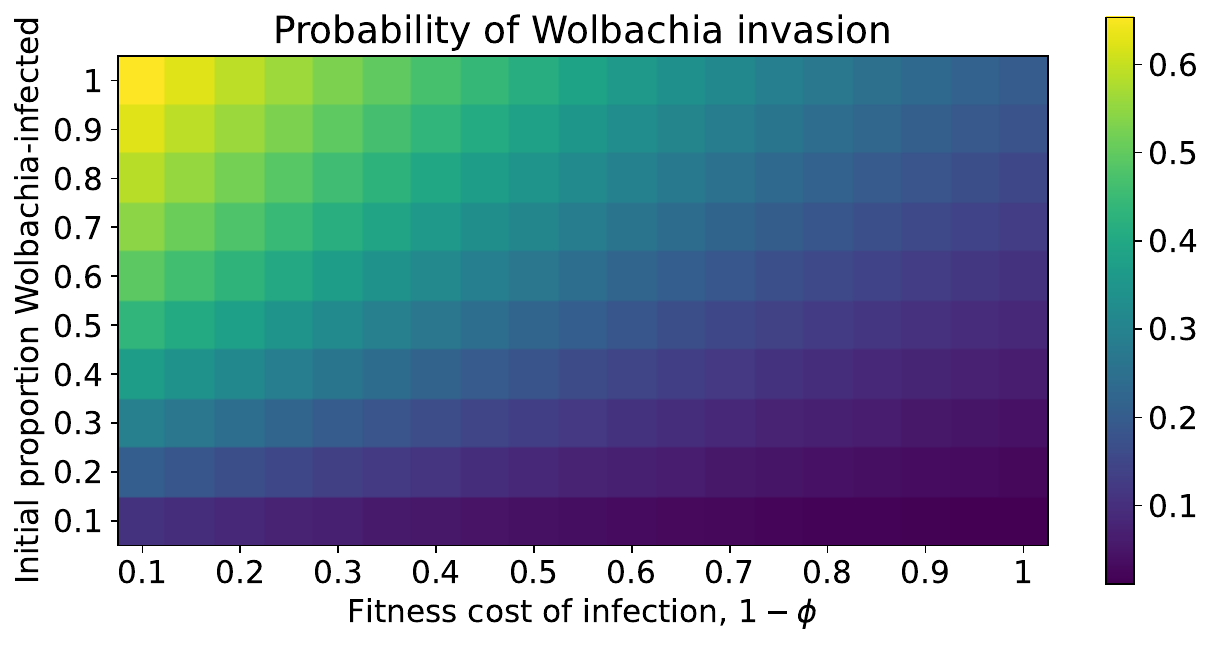}
    \caption{Probability of successful Wolbachia invasion of a household depending on the initial (female) Wolbachia-infected proportion $N_w/N_0$ and fitness cost of infection $1-\phi$. Initially the household is composed of $N_0=10$ female wildtype mosquitoes and between $1$ and $10$ additional Wolbachia-infected females are introduced so the total household population is between $11$ and $20$. Households can have up to $30$ female mosquitoes and $v=0.9$ and the larval density function is \eqref{eq:hughes_F}.}
    \label{fig:color_plot_invasion_prob_Dye_rev}
\end{figure}

\begin{figure}
    \centering
    \includegraphics[width=0.7\linewidth]{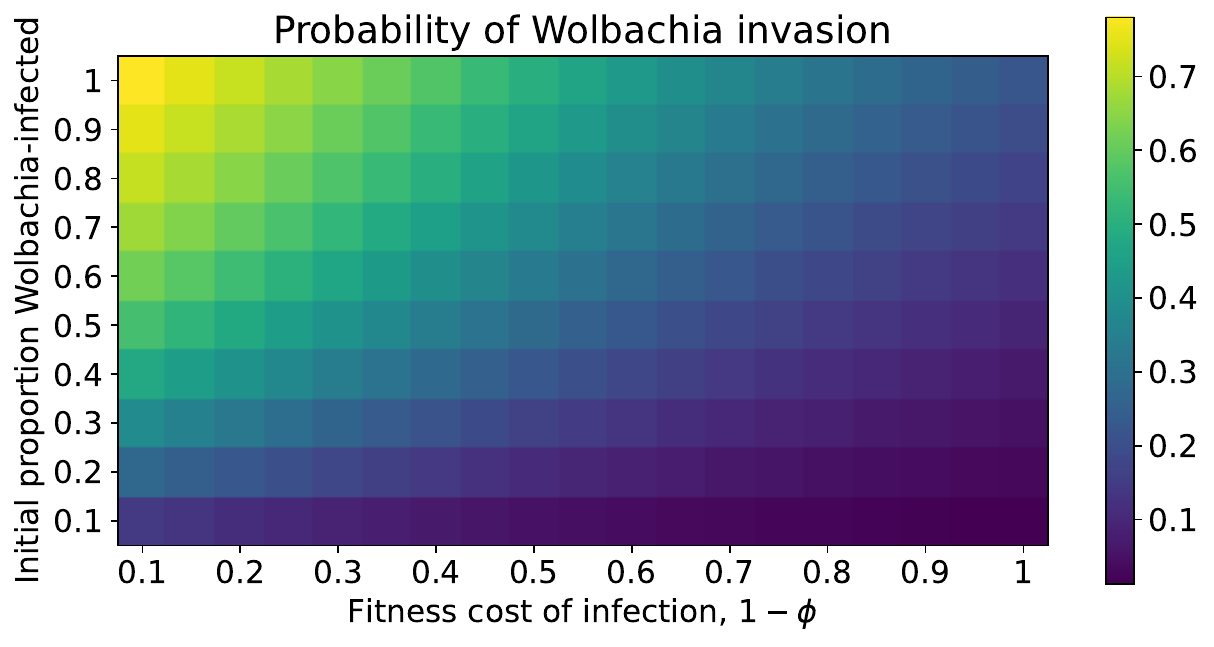}
    \caption{Probability of successful Wolbachia invasion of a household depending on the initial (female) Wolbachia-infected proportion $N_w/N_0$ and fitness cost of infection $1-\phi$. Initially the household is composed of $N_0=10$ female wildtype mosquitoes and between $1$ and $10$ additional Wolbachia-infected females are introduced so the total household population is between $11$ and $20$. Households can have up to $30$ female mosquitoes and $v=1$ and the larval density function is \eqref{eq:qu}.}
    \label{fig:qu_v1}
\end{figure}

\begin{figure}
    \centering
    \includegraphics[width=0.7\linewidth]{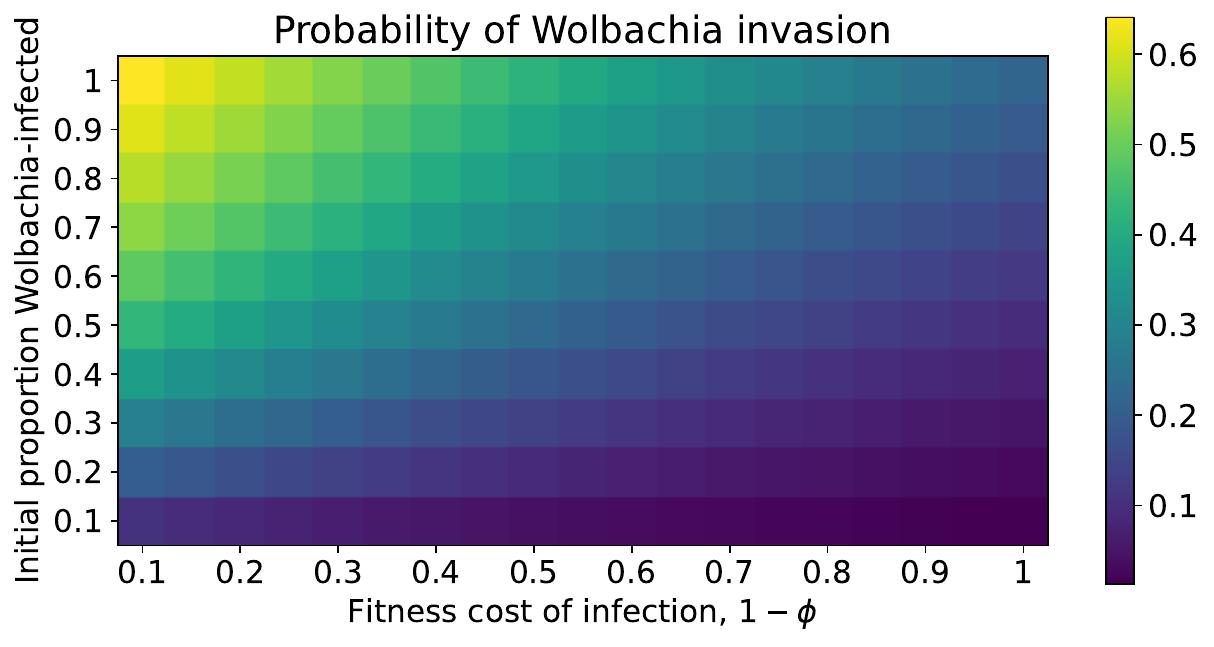}
    \caption{Probability of successful Wolbachia invasion of a household depending on the initial (female) Wolbachia-infected proportion $N_w/N_0$ and fitness cost of infection $1-\phi$. Initially the household is composed of $N_0=10$ female wildtype mosquitoes and between $1$ and $10$ additional Wolbachia-infected females are introduced so the total household population is between $11$ and $20$. Households can have up to $30$ female mosquitoes and $v=0.9$ and the larval density function is \eqref{eq:qu}.}
    \label{fig:qu_v09}
\end{figure}

In this section, we show plots for the probability of Wolbachia invasion into a wildtype-only mosquito household currently at its deterministic steady state value, $m^* = 10$. We vary the number of Wolbachia-infected mosquitoes released as a proportion of the initial wildtype household size $N_m(0)$, and add these to the initial wildtype population rather than replacing the corresponding number of wildtype mosquitoes. We vary the fitness cost of infection $1-\phi$ between $0.1$ and $1$. Note that here we are allowing the total number of mosquitoes in the household to be pushed above the steady state value. The plots include both larval density dependence functions \eqref{eq:hughes_F} (see Figures \ref{fig:hughes_v1} and \ref{fig:color_plot_invasion_prob_Dye_rev}) and \eqref{eq:qu} (see Figures \ref{fig:qu_v1} and \ref{fig:qu_v09}) and the cases where $v=1$ and $v=0.9$.


\section{Wolbachia invasion results, $v=0.9$}\label{ap:reversion}
\begin{figure}
    \centering
    \includegraphics[width=0.7\linewidth]{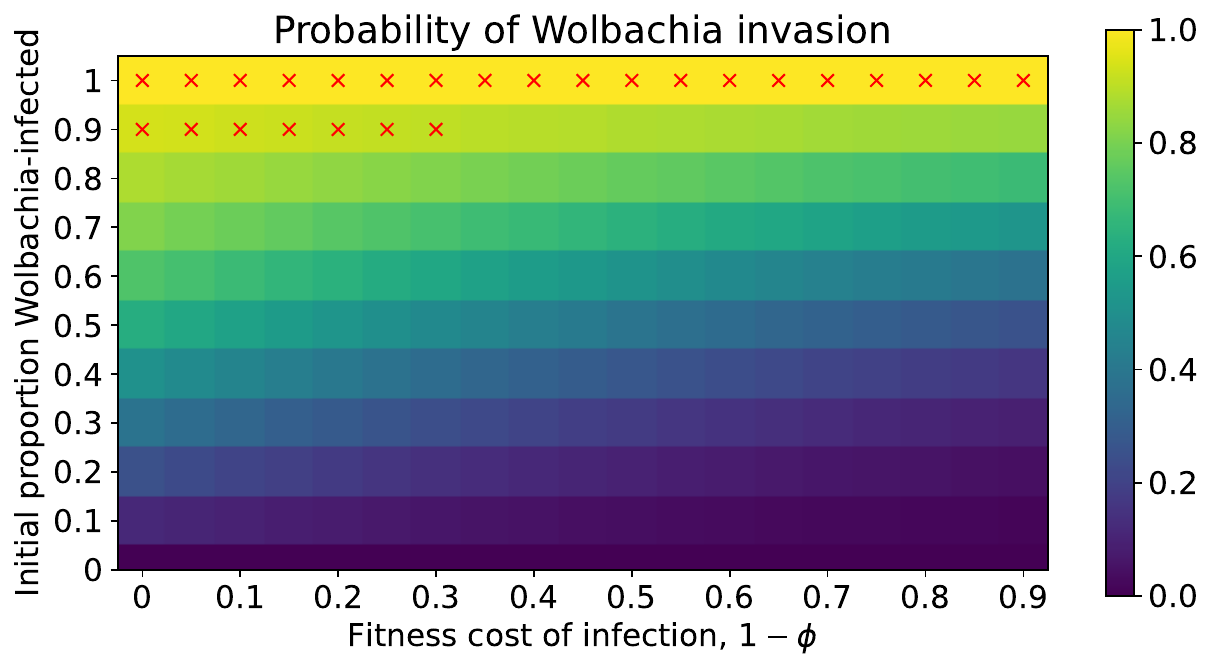}
    \caption{Probability of successful Wolbachia invasion of a household depending on the initial Wolbachia-infected proportion $N_w/N_0$, fitness cost of infection $1-\phi$, $v=0.9$ with larval density function \eqref{eq:hughes_F}. The initial household size is fixed at $N_0=10$ female mosquitoes and the initial infected proportion is determined by $N_w(0)$ with $N_m(0) + N_w(0) = N_0$. Households can have up to $30$ female mosquitoes. The red crosses indicate where the probability of successful Wolbachia invasion is greater than or equal to $0.9$.}
    \label{fig:color_stab_v09_dye}
\end{figure}

In this section we present results on the probability of Wolbachia invasion and the expected time until invasion consistent with those from Section \ref{sec:30mosq}, except $v=0.9$ so that reversion is possible. Here we use the original larval density function $F(N)$ from the main text \eqref{eq:hughes_F}. The results are similar to when $v=1$. However, observe the increase in proportion of Wolbachia-infected mosquitoes required in a household of size $10$ to achieve a successful invasion with probability $0.9$ in Figure \ref{fig:color_stab_v09_dye} (compare with Figure \ref{fig:color_stab_dye_v1}).

\begin{figure}
  \centering
  \begin{subfigure}{0.4\textwidth}
    \includegraphics[width=\linewidth]{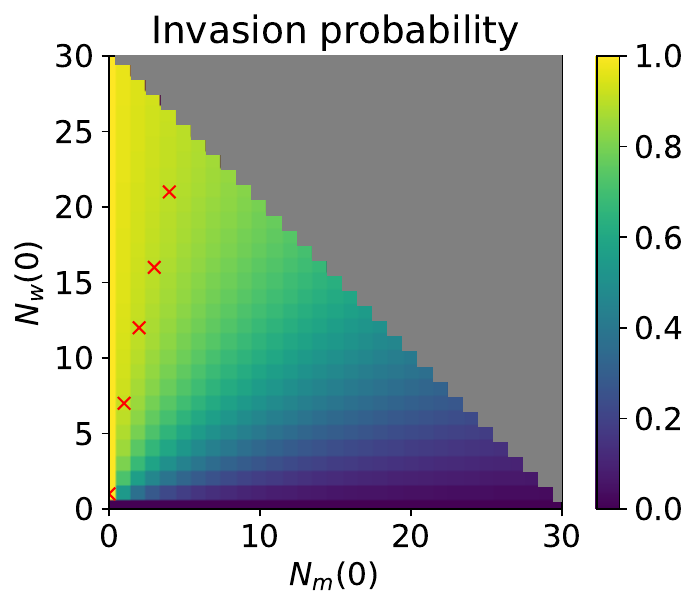}
    \caption{}
    \label{fig:subfig1_prob_invade_v}
  \end{subfigure}
  \hfill
  \begin{subfigure}{0.4\textwidth}
    \includegraphics[width=\linewidth]{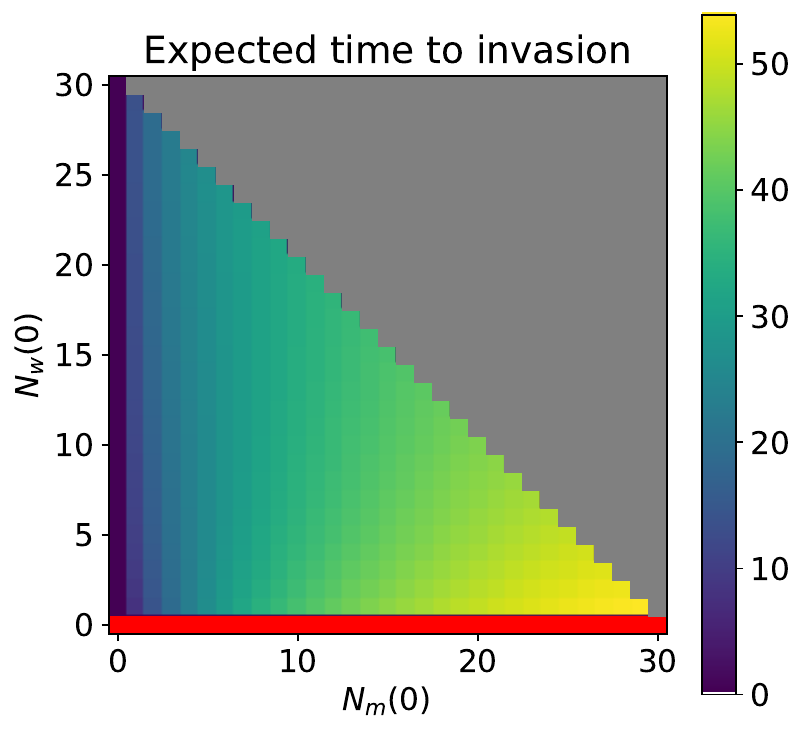}
    \caption{}
    \label{fig:subfig2_time_invade_v}
  \end{subfigure}
  \caption{(a) Probability of entering the Wolbachia-infected class when the initial state is $(N_m(0), N_w(0))$. Red crosses denote, for different initial wildtype mosquito numbers, the minimum number of Wolbachia-infected mosquitoes required for successful invasion with probability $0.9$ or higher. Note that, if there are more than $4$ wildtype female mosquitoes present, the number of Wolbachia-infected mosquitoes required for a successful invasion with probability $0.9$ or higher is outside of the state space. (b) Expected time in days to enter the Wolbachia-infected class, conditional on this event occurring, when the initial state is $(N_m(0), N_w(0))$. The red bar indicates initial conditions where only wildtype mosquitoes are present and so the Wolbachia-infected class cannot be reached. In both diagrams the maximum number of female mosquitoes in a household is $30$, the initial state probability distribution is $P_{(N_m(0),N_w(0))}(0)=1$, parameters are as in Table \ref{tab:params}, $1-\phi=0.15$, $v=0.9$, and the larval density function is \eqref{eq:hughes_F}. The top right quadrant shaded grey is not an admissible region of the state space.}\label{fig:color_plots_dye_v09}
  \end{figure}

\begin{figure}
    \centering
    \includegraphics[width=0.5\linewidth]{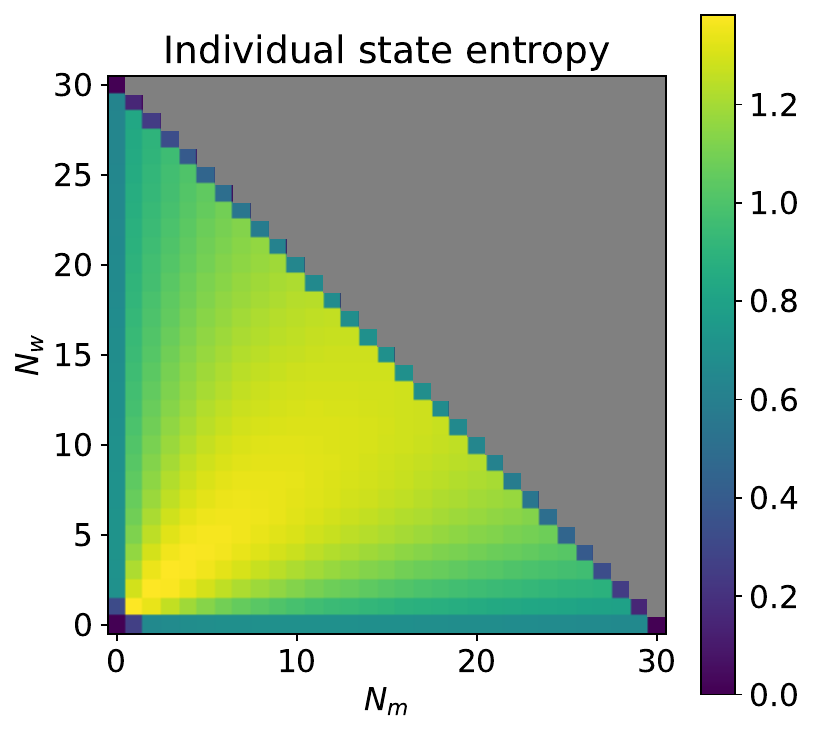}
    \caption{Entropy of state $(N_m,N_w)$. Higher entropy corresponds to greater uncertainty in the next state. The maximum number of female mosquitoes in household is $30$, parameters are as in Table \ref{tab:params}, $1-\phi=0.15$, $v=0.9$, larval density function is \eqref{eq:qu}. The top right quadrant shaded grey is not an admissible region of the state space.}
    \label{fig:entropies_hughes_Dye_v}
\end{figure}

In Figure \ref{fig:subfig1_prob_invade_v} the minimum number of Wolbachia-infected mosquitoes required for the probability of invasion to be at least $0.9$ has dramatically increased (as well as all the invasion probabilities across the initial condition space) compared to Figure \ref{fig:color_plots_dye_v1}. This is because the Wolbachia-infected birth rate is now leaky. The expected time until Wolbachia invasion in Figure \ref{fig:subfig2_time_invade_v} remains comparable to Figure \ref{fig:subfig2_time_invade} as well as the state entropies in Figure \ref{fig:entropies_hughes_Dye_v} with previous Figure \ref{fig:entropies_hughes_Dye}.

\section{Calculating the time until Wolbachia extinction}\label{ap:extinction_time}
Here, we use the same Markov chain theory as in Section \ref{subsec:absorbtime} \cite{norris1998markov,gomez2012extinction} to determine the expected time until extinction of a Wolbachia-only population when $v=1$ so that reversion cannot occur. Recall $\bm{Q}_2$ is the sub-$Q$ matrix corresponding to the Wolbachia-only class and the probability of eventually entering the absorbing state $(0,0)$ from any state in $\mcs{}$ is $1$. The quantity $\mathbf{q}(0,0)$ is the vector of transition rates from each state in $\mcs{2}$ to the absorbing state.
The expected time until extinction of a Wolbachia-only household in state $(0,w')$ is the solution of the linear system,
\begin{align}
    \sum_{(0,l)\in\{(0,0)\}\cup \mcs{2}} q_{(0,w'),(0,l)}\tau_{(0,l)}(0,0)=-1\;\;\; \text{for}\;\;\; (0,w')\in \mcs{2},
\end{align}
where $q_{(0,w'),(0,l)}$ is the transition rate from state $(0,w')$ to state $(0,l)\in \{(0,0)\}\cup \mcs{2}$ and $\tau_{(0,l)}(0,0)$ is the expected time for the household to enter the absorbing state $(0,0)$ from state $(0,l)$.

\section{The Wolbachia-only QSD} \label{ap:wolb_QSD}
\begin{figure}
    \centering
    \includegraphics[width=0.5\linewidth]{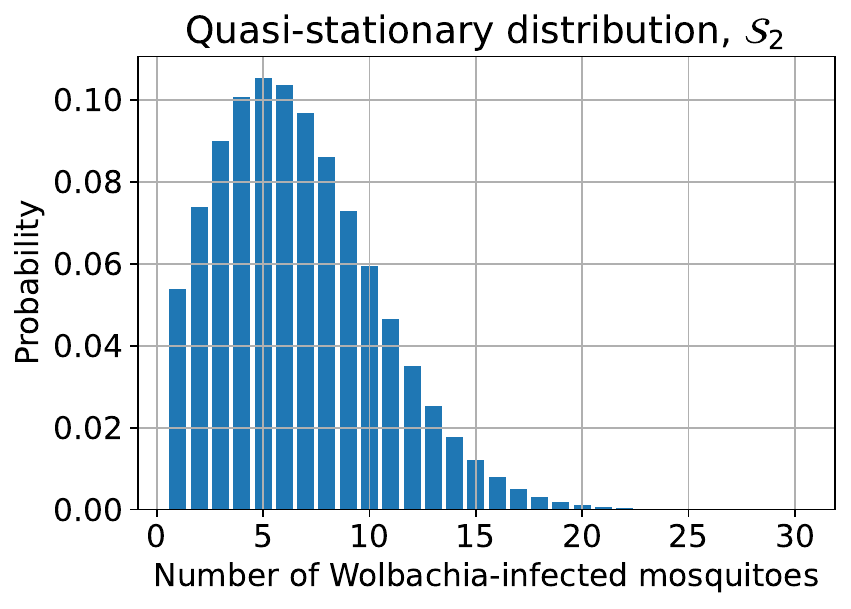}
    \caption{
    QSD for the Wolbachia-only class for the model with households of up to $30$ female mosquitoes, perfect vertical transmission ($v=1$), $1-\phi=0.15$ and larval density function \ref{eq:hughes_F}. Any states not shown have probability close to $0$.}
    \label{fig:qsd_wolb}
\end{figure}

The QSD for the Wolbachia-only class is found by setting $v=1$ under the $30$ mosquito CTMC model. This ensures that the wildtype-only class is not accessible via the Wolbachia-only class. We then find the minimal magnitude eigenvalue for the Wolbachia-only class. This corresponds to the minimal magnitude eigenvalue of $\bm{Q}_2$. Then the Wolbachia-only QSD is given by the left eigenvector of $\bm{Q}$ associated with that eigenvalue, normalised to sum to one. The Wolbachia-only QSD is shown in Figure \ref{fig:qsd_wolb}.

\section{Deriving the damping ratio for a CTMC model}\label{ap:damping_ratio}
We can omit the absorbing state $(0,0)$ and its corresponding eigenvalue from equation \eqref{eq:cnt_eigs} in order to consider just the transient subsystem \cite{mubayi2019studying}. Let $-\lambda_1=-\alpha$, be the minimal magnitude eigenvalue and let $-\lambda_2$ be the eigenvalue of second smallest magnitude. Consequently, we can write
\begin{align}
    \bm{P}(t)=\sum_{i=1}^{s-1}c_i e^{-\lambda_it}\bm{l}_i,
\end{align}
and if we divide equation \eqref{eq:cnt_eigs} by $\exp{(-\lambda_1 t)}$ we get
\begin{align}
    \frac{\bm{P}(t)}{e^{-\lambda_1 t}} = c_1\bm{l}_1 + c_2\left(\frac{e^{-\lambda_2}}{e^{-\lambda_1}}\right)^t \bm{l}_2 + \ldots,
\end{align}
and taking the limit, 
\begin{align}
 \lim_{t \rightarrow \infty} \frac{\bm{P}(t)}{e^{-\lambda_1 t}}= c_1\bm{l}_1.
\end{align}
Therefore, it follows that \cite{caswell2000matrix} 
\begin{align}
    \lim_{t \rightarrow \infty} \left(\frac{\bm{P}(t)}{e^{-\lambda_1 t}}-c_1\bm{l}_1 - c_2\rho^{-t}\bm{l}_2\right)=0,
\end{align}
where the damping ratio $\rho$ is defined as
\begin{align}
    \rho=\exp((-\lambda_1)-(-\lambda_2))=\exp(\lambda_2-\lambda_1).
\end{align}
\section{Deriving the expected time until Wolbachia invasion}\label{ap:absorb_time}
Here, we derive the expected time $\tau^*_{(k,l)}$ starting from state $(k,l) \in \mcs{3}\cup \mcs{2}$ until communicating class $\mcs{2}$ is hit, given that $\mcs{1}$ is not entered first. Adapted from \cite{gomez2012extinction}, $\tau^*_{(k,l)}$ is the minimal solution of \eqref{eq:absrobtime} in Section \ref{subsec:absorbtime}.
 
To show this here, we relabel the states with a single letter for ease of notation. For $i\in \mcs{3}\cup \mcs{2}$, we can write the expected time to hit communicating class $\mcs{2}$ given that the starting state is $i$ and that $\mcs{1}$ is not reached first,
\begin{align}
    \tau_i^*=\mathbb{E}[T|X(0)=i,X(T)\in\mcs{2}],
\end{align}
where $\mathcal{X}:=\{X(t), t\geq 0\}$ is the Markov chain and $T=\inf\{t\geq 0:X(t)\in \mcs{1}\cup\mcs{2}\}$ is the first time we leave the mixed class. Now, by the conditional expectation formula,
\begin{align}
    \tau_i(w)=\frac{\mathbb{E}_i\left[T\mathbbm{1}_{\{X(T)\in \mcs{2}\}}\right]}{\mathbb{P}_i(X(T)\in \mcs{2})},
\end{align}
where we use the standard notation $\mathbb{P}_i(\cdot)=\mathbb{P}(\cdot | X(0)=i)$ and $\mathbb{E}_i(\cdot)=\mathbb{E}[\cdot |X(0)=i]$. Let $a_i^*=\mathbb{P}_i(X(T)\in \mcs{2})$; then for $i\in\mcs{3}$, letting $J_1$ denote the first jump time of $\mathcal{X}$ and letting $\pi_{ij}=\mathbb{P}(X(J_1)=j|X(0)=i)$ denote an entry of the jump matrix associated with $\bm{Q}$, we can write
\begin{equation}
\begin{aligned}
    \tau_i^*a_i^* &= \mathbb{E}_i\left[T\mathbbm{1}_{\{X(T)\in\mcs{2}\}}\right]\\
    &=\mathbb{E}_i\left[(J_1+ (T-J_1))\mathbbm{1}_{\{X(T)\in \mcs{2}\}}\right]\\
    & =\mathbb{E}_i\left[J_1\mathbbm{1}_{\{X(T)\in \mcs{2}\}}\right]+\sum_{j\neq i, \;j\in \mcs{3}\cup\mcs{2}} \pi_{ij}\mathbb{E}_j\left[T\mathbbm{1}_{\{X(T)\in \mcs{2}\}}\right].
\end{aligned}
\end{equation}
Where the last line follows by the strong Markov property at time $J_1$. Further, since $J_1$ and $\{X(t+J_1),t\geq 0\}$ are independent and since $\pi_{ij}=q_{ij}/q_i$ by the jump matrix definition \cite{norris1998markov}, we can write
\begin{equation}
\begin{aligned}
\tau_i^*a_i^*&=\mathbb{E}_i[J_1]\mathbb{E}_i\left[\mathbbm{1}_{\{X(T)\in\mcs{2}\}}\right]+\sum_{j\neq i,\;j\in \mcs{3}\cup\mcs{2}}\frac{q_{ij}}{q_i}\tau_j^*a_j^*\\
&=q^{-1}_i\mathbb{P}_i(X(T)\in \mcs{2})+\sum_{j\neq i,\;j\in \mcs{3}\cup \mcs{2}} \frac{q_{ij}}{q_i}\tau_j^*a_j^*\\
&=q^{-1}_ia_i^* + \sum_{j\neq i,\;j\in \mcs{3}\cup\mcs{2}} \frac{q_{ij}}{q_i}\tau_j^*a_j^*,
\end{aligned}
\end{equation}
where the second line follows since, conditional on $X(0)=i$, $J_1$ is exponentially distributed with rate $q_i$. Therefore,
\begin{equation}
\begin{aligned}
    &\tau_i^*a_i^*q_i=a_i^*+\sum_{j\neq i,\;j\in \mcs{3}\cup\mcs{2}} q_{ij}\tau_j^*a_j^*\\
    &\Rightarrow \sum_{j\in \mcs{3}\cup\mcs{2}} q_{ij}\tau_j^*a_j^*=-a_i^*,
\end{aligned}
\end{equation}
since $q_i=-q_{ii}$. This yields the required result \eqref{eq:absrobtime} in the compact notation.

\section{Eigenvalue interpretation justification}\label{ap:proofs}
\subsection{Claim 1: Time for probability mass in class to become negligible}\label{ap:claim1}
We claimed that the time for the probability mass in class $\mcs{k}$ in our system to become negligible is governed by the minimal magnitude eigenvalue of $\bm{Q}_k$. We will now prove this.

Let $\bm{Q}_k$ denote the matrix $\bm{Q}$ restricted to $\mcs{k}$ and let $a_k=\max_{i\in \mcs{k}}(-\bm{Q}_{ii})>0$, the maximal magnitude of the diagonal entries of $\bm{Q}_k$. Since $\mcs{k}$ is a communicating class, for all $i,j\in \mcs{k}$, there exist an $n \in \mathbb{N}$ and $i=i_0,i_1,\ldots ,i_n=j\in \mcs{k}$ such that $(\bm{Q}_k)_{i_r,i_r+1}>0$ for all $r \in \{0,\ldots ,n-1\}$. Therefore $\bm{Q}_k+a_kI$ is a non-negative irreducible matrix. Hence we can apply the Perron-Frobenius Theorem to $\bm{Q}_k + a_kI$. This tells us that there exists a unique left eigenvector $\bm{l}$ of the matrix $\bm{Q}_k + a_kI$ such that each entry of $\bm{l}$ is positive, and a unique right eigenvector $\bm{u}$ of the matrix $\bm{Q}_k + a_kI$ such that each entry of $\bm{u}$ is positive, both with the same positive eigenvalue $b_k$. The eigenvalue $b_k$ is the largest eigenvalue of $\bm{Q}_k + a_kI$, and is simple. Letting $-\alpha_k=b_k-a_k$, we have that $\bm{l}\bm{Q}_k=-\alpha_k\bm{l}$ and $\bm{Q}_k\bm{u}=-\alpha_k\bm{u}$, and $-\alpha_k$ is the largest eigenvalue of $\bm{Q}_k$ and is simple. Therefore, there exists an invertible matrix $\bm{A}_k$ such that $\bm{Q}_k=\bm{A}_k\bm{\Lambda}_k\bm{A}^{-1}_k$, where
\begin{align}\label{eq:lambda_k}
    \bm{\Lambda}_k=\begin{pmatrix}
        -\alpha_k & \bm{0}_{1\times n} \\
        \bm{0}_{n\times 1} & \bm{\Lambda}'_k \\
    \end{pmatrix},
\end{align}
$n=|\mcs{k}|-1$ and $\bm{0}_{n\times 1}=\bm{0}_{1\times n}^{\top}$ is a column vector of zeros and $\bm{\Lambda}'_k$ is a Jordan normal matrix with diagonal entries strictly less than $-\alpha_k$. Since $(\bm{Q}_k\bm{A}_k)_{i1}=\sum_{j\in\mcs{k}}(\bm{Q}_k)_{ij}(\bm{A}_k)_{j1}=(\bm{A}_k\bm{\Lambda}_k)_{i1}=-\alpha_k(\bm{A}_k)_{i1}$, it holds that $(\bm{A}_k)_{i1}=\tilde{c}\bm{u}_i$ for some constant $\tilde{c}$. Since $(\bm{A}^{-1}_k\bm{Q}_k)_{1i}=\sum_{j\in\mcs{k}}(\bm{A}^{-1}_k)_{1j}(\bm{Q}_k)_{ji}=(\bm{\Lambda}_k\bm{A}^{-1}_k)_{1i}=-\alpha_k(\bm{A}^{-1}_k)_{1i}$, we have also that $(\bm{A}^{-1}_k)_{1i}=c\bm{l}_i$ for some constant $c$. Since $\sum_{i \in \mcs{k}}(\bm{A}_k^{-1})_{1i}(\bm{A}_k)_{i1}=1$, we have $\sum_{i\in \mcs{k}}c\tilde{c}\bm{l}_i\bm{u}_i=1$. Therefore we can assume that $\bm{u}$ and $\bm{l}$ are chosen in such a way that $(\bm{A}_k)_{i1}=\bm{u}_i$, $(\bm{A}_k^{-1})_{1i}=\bm{l}_i$ and $\bm{l}\cdot \bm{u}=1$.

For $i,j\in \mcs{k}$ and $t\geq 0$, we have \cite{norris1998markov}
\begin{align}
    \mathbb{P}(X(t)=j|X(0)=i)=(e^{t\bm{Q}})_{ij} = \sum_{m=0}^{\infty}\frac{1}{m!}t^m(\bm{Q}^m)_{ij}=\sum_{m=0}^{\infty}\frac{1}{m!}t^m(\bm{Q}^m_k)_{ij},
\end{align}
since, by the class structure of the Markov chain, for $n\in \mathbb{N}$ and $i=i_0,i_1,\ldots ,i_n=j\in \mathcal{S}$ with $i,j\in\mcs{k}$ and $i_{n'}\notin \mcs{k}$ for some $n'\in \{1,\ldots ,n-1\}$ we have $\bm{Q}_{i_0,i_1}\bm{Q}_{i_1,i_2}\ldots \bm{Q}_{i_{n-1},i_n}=0$. Therefore
\begin{equation}\label{eq:H20}
\begin{aligned}
    \mathbb{P}(X(t)=j|X(0)=i)&=\sum_{m=0}^{\infty}\frac{1}{m!}t^m\left(\bm{A}_k\bm{\Lambda}_k\bm{A}^{-1}_k\right)^m_{ij}=\sum_{m=0}^{\infty}\frac{1}{m!}t^m\left(\bm{A}_k\bm{\Lambda}_k^m\bm{A}^{-1}_k\right)_{ij}\\
    &=\left(\bm{A}_k\left(\sum_{m=0}^{\infty}\frac{1}{m!}t^m\bm{\Lambda}^m_k\right)\bm{A}^{-1}_k\right)_{ij}.
\end{aligned}
\end{equation}
By \eqref{eq:lambda_k}, we also have that
\begin{align}
    \sum_{m=0}^{\infty}\frac{1}{m!}t^m\bm{\Lambda}^m_k=\begin{pmatrix}
        e^{-\alpha_kt} & \bm{0}_{1\times n} \\
        \bm{0}_{n \times 1} & \bm{B}_k(t)
    \end{pmatrix},
\end{align}
where the entries of $\bm{B}_k(t)$ are all at most $t^{|\mcs{k}|-2}e^{-\alpha'_k t}$, where $-\alpha'_k<-\alpha_k$ is the second largest eigenvalue of $\bm{Q}_k$. So substituting in \eqref{eq:H20}, it holds that as $t \rightarrow \infty$
\begin{equation}\label{eq:H22}
\begin{aligned}
    \mathbb{P}(X(t)=j|X(0)=i)&=(\bm{A}_k)_{i1}e^{-\alpha_kt}\left(\bm{A}^{-1}_k\right)_{1j} +O\left(t^{|\mcs{k}|-2}e^{-\alpha'_kt}\right)\\
    &=e^{-\alpha_kt}\bm{u}_i\bm{l}_j+O\left(t^{|\mcs{k}|-2}e^{-\alpha'_kt}\right),
\end{aligned}
\end{equation}
where eigenvectors $\bm{l}$ and $\bm{u}$ are both positive and satisfy $\bm{l}\bm{Q}_k=-\alpha_k\bm{l}$, $\bm{Q}_k\bm{u}=-\alpha_k\bm{u}$ and $\bm{l} \cdot \bm{u}=1$, and $-\alpha_k$ is the largest eigenvalue of $\bm{Q}_k$. This is the required result.

\subsection{Claim 2: Quasi-stationary distribution with a class}\label{ap:claim2}
Secondly, in Section \ref{subsec:com-class} we claimed that the probability dynamics in our system in $\mcs{2}$ will enter their own kind of quasi-stationary distribution and that the same holds for $\mcs{1}$ (which is trivial since the QSD of the full system only holds positive probability mass in $\mcs{1}$). We prove this claim next below.

Fix $k\in \{1,2\}$ and $j \in \mcs{k}$, then the probability that the Markov chain is at state $j$ at time $t$, given that it started at state $i\in \mcs{3}$ and given that it is in $\mcs{k}$ at time $t$ is
\begin{align}\label{eq:H23}
    \mathbb{P}(X(t)=j|X(t)\in\mcs{k},X(0)=i)=\frac{\mathbb{P}(X(t)=j|X(0)=i)}{\mathbb{P}(X(t)\in \mcs{k}|X(0)=i)}.
\end{align}
Our strategy will be to show that given $X(0)=i$, the events $\{X(t)=j\}$ and $\{X(t)\in \mcs{k}\}$ are very unlikely to occur unless the Markov chain is already in class $\mcs{k}$ at an earlier time $ct$, for suitable $c<1$. Then between the times $ct$ and $t$, the Markov chain can get close to the QSD in $\mcs{k}$.

Recall from Section \ref{subsec:30mosq_results} that $\alpha_3>\alpha_k$ (see first row of Table \ref{tab:eigs}). So we can take $c<1$ sufficiently large that $\alpha_3c>\alpha_k$. Then it holds that
\begin{equation}\label{eq:H24}
\begin{aligned}
    \mathbb{P}(X(t)=j|X(0)=i)&=\mathbb{P}(X(t)=j, X(ct)\in \mcs{k}|X(0)=i) \\&+ \mathbb{P}(X(t)=j,X(ct)\in\mcs{3}|X(0)=i).
\end{aligned}
\end{equation}
That is, the probability the Markov chain is at state $j$ at time $t$ (given it started at state $i$) can be partitioned into the sum of the probabilities that the Markov chain has already entered $\mcs{k}$ at time $ct$ or that the Markov chain is still in $\mcs{3}$ at time $ct$. Clearly, 
\begin{equation}
\begin{aligned}
        \mathbb{P}(X(t)=j,X(ct)\in\mcs{3}|X(0)=i)&\leq \mathbb{P}(X(ct)\in \mcs{3}|X(0)=i)\\
        &=\sum_{j'\in \mcs{3}} \mathbb{P}(X(ct)=j'|X(0)=i)=O\left( e^{-\alpha_3 ct}\right),
\end{aligned}
\end{equation}
where the last line follows by \eqref{eq:H22} because $\mathbb{P}(X(ct)=j'|X(0)=i)=O\left(e^{-\alpha_3ct}\right)$ for each $j'\in\mcs{3}$.  For the first term on the right hand side of \eqref{eq:H24}, since if $X(1)\in\mcs{k}$ and $X(t)\in\mcs{k}$ then $X(ct)\in \mcs{k}$, we can fix $j_0\in\mcs{k}$ and write 
\begin{equation}
    \begin{aligned}
        \mathbb{P}(X(t)=j,X(ct)&\in\mcs{k}|X(0)=i)\geq \mathbb{P}(X(t)=j,X(1)=j_0|X(0)=i)\\
        &=\mathbb{P}(X(t-1)=j|X(0)=j_0)\mathbb{P}(X(1)=j_0|X(0)=i),
    \end{aligned}
\end{equation}
where the last line follows by the Markov property at time $1$. By \eqref{eq:H22}, and since $j,j_0 \in \mcs{k}$, we have $\mathbb{P}(X(t-1)=j|X(0)=j_0) \gtrsim e^{-\alpha_k(t-1)}$ as $t\rightarrow\infty$.

Since $\mathbb{P}(X(1)=j_0|X(0)=i)>0$ because $i \in \mcs{3}$ and $j_0\in \mcs{k}$ and so $i\rightarrow j_0$, it follows that $\mathbb{P}(X(t)=j,X(ct)\in\mcs{k}|X(0)=i) \gtrsim e^{-\alpha_kt}$ as $t\rightarrow \infty$. Therefore,
\begin{equation}\label{eq:*}
    \begin{aligned}
        \frac{\mathbb{P}\left(X(t)=j,X(ct)\in\mcs{3}|X(0)=i\right)}{\mathbb{P}\left(X(t)=j,X(ct)\in\mcs{k}|X(0)=i\right)}&=O\left(e^{-\alpha_3ct+\alpha_kt}\right)\\
        &=O\left(e^{-c't}\right),
    \end{aligned}
\end{equation}
where $c'=\alpha_3c-\alpha_k>0$ by our choice of $c$. By \eqref{eq:H24} and \eqref{eq:*} we now have \begin{equation}\label{eq:**}
     \begin{aligned}
         \mathbb{P}(X(t)=j|X(0)=i)=\left(1+\epsilon_t^{(j)}\right)\mathbb{P}\left(X(t)=j,X(ct)\in\mcs{k}|X(0)=i\right),
     \end{aligned}
 \end{equation}
where $\epsilon_t^{(j)}=O\left(e^{-c't}\right)$ as $t \rightarrow \infty$. Since \eqref{eq:**} holds for each $j\in\mcs{k}$, we can also write
\begin{equation}\label{eq:cross}
    \begin{aligned}
        \mathbb{P}(X(t)\in\mcs{k}|X(0)=i)&=\sum_{j'\in\mcs{k}}\mathbb{P}(X(t)=j'|X(0)=i)\\
        &=\sum_{j'\in\mcs{k}}\left(1+\epsilon_t^{(j')}\right)\mathbb{P}(X(t)=j', X(ct)\in \mcs{k}|X(0)=i)\\
        &\leq \left(1+\max_{j'\in\mcs{k}}\epsilon_t^{(j')}\right)\sum_{j'\in\mcs{k}}\mathbb{P}(X(t)=j', X(ct)\in \mcs{k}|X(0)=i)\\
        &=\left(1+\max_{j'\in\mcs{k}}\epsilon_t^{(j')}\right)\mathbb{P}(X(t)\in\mcs{k}, X(ct)\in \mcs{k}|X(0)=i).
    \end{aligned}
\end{equation}
Since we also have $\mathbb{P}(X(t)\in\mcs{k}|X(0)=i)\geq\mathbb{P}(X(t)\in \mcs{k}, X(ct)\in\mcs{k}|X(0)=i)$, it follows from \eqref{eq:cross} that for some $\epsilon'_t\geq0$ with $\epsilon'_t\leq \max_{j'\in\mcs{k}}\epsilon^{(j')}_t$, we have
\begin{equation}\label{eq:star}
    \begin{aligned}
        \mathbb{P}(X(t)\in\mcs{k}|X(0)=i)=\left(1+\epsilon'_t\right)\mathbb{P}(X(t)\in\mcs{k},X(ct)\in\mcs{k}|X(0)=i).
    \end{aligned}
\end{equation}
Substituting in \eqref{eq:**} and \eqref{eq:star} into \eqref{eq:H23}, we now have
\begin{equation}\label{eq:2star}
\begin{aligned}
    &\mathbb{P}(X(t)=j|X(t)\in\mcs{k},X(0)=i)
    =\frac{\left(1+\epsilon_t^{(j)}\right)\mathbb{P}(X(t)=j, X(ct)\in \mcs{k}|X(0)=i)}{\left(1+\epsilon'_t\right)\mathbb{P}(X(t)\in\mcs{k},X(ct)\in\mcs{k}|X(0)=i)}\\
    &=\frac{1+\epsilon_t^{(j)}}{1+\epsilon'_t}\frac{\sum_{j'\in\mcs{k}}\mathbb{P}(X(t)=j,X(ct)=j'|X(0)=i)}{\sum_{j'\in\mcs{k}}\mathbb{P}(X(t)\in \mcs{k},X(ct)=j'|X(0)=i)}\\
    &=\frac{1+\epsilon_t^{(j)}}{1+\epsilon'_t}\frac{\sum_{j'\in\mcs{k}}\mathbb{P}(X(ct)=j'|X(0)=i)\mathbb{P}(X((1-c)t)=j|X(0)=j')}{\sum_{j'\in\mcs{k}}\mathbb{P}(X(ct)=j'|X(0)=i)\mathbb{P}(X((1-c)t)\in\mcs{k}|X(0)=j')},
\end{aligned}
\end{equation}
where the last line follows by the Markov property at time $ct$. For $j'\in\mcs{k}$, by \eqref{eq:H22} we have
\begin{equation}
    \begin{aligned}
        \frac{\mathbb{P}(X((1-c)t)=j|X(0)=j')}{\mathbb{P}(X((1-c)t)\in\mcs{k}|X(0)=j')}=\frac{\mathbb{P}(X((1-c)t)=j|X(0)=j')}{\sum_{j''\in\mcs{k}}\mathbb{P}(X((1-c)t)=j''|X(0)=j')}\\
        =\frac{e^{-\alpha_k(1-c)t}\bm{u}_{j'}^{(k)}\bm{l}_j^{(k)}+O\left(t^{|\mcs{k}|-2}e^{-\alpha'_k(1-c)t}\right)}{\sum_{j''\in\mcs{k}}\left(e^{-\alpha_k(1-c)t}\bm{u}_{j'}^{(k)}\bm{l}_{j''}^{(k)}\right)+O\left(t^{|\mcs{k}|-2}e^{-\alpha'_k(1-c)t}\right)},
    \end{aligned}
\end{equation}
where $\bm{l}^{(k)}$ and $\bm{u}^{(k)}$ are the positive eigenvectors of $\bm{Q}_k$ with $\bm{l}^{(k)}\bm{Q}_k=-\alpha_k\bm{l}^{(k)}$, $\bm{Q}_k\bm{u}^{(k)}=-\alpha_k\bm{u}^{(k)}$ and $\bm{l}^{(k)} \cdot \bm{u}^{(k)}=1$ and $-\alpha_k'<-\alpha_k$ is the second largest eigenvalue of $\bm{Q}_k$. Therefore, since the $\bm{u}^{(k)}_{j'}$ terms and $e^{-\alpha_k(1-c)t}$ terms cancel, we have
\begin{equation}
    \begin{aligned}
        \frac{\mathbb{P}(X((1-c)t)=j|X(0)=j')}{\mathbb{P}(X((1-c)t)\in\mcs{k}|X(0)=j')}=\frac{\bm{l}_j^{(k)}+O\left(t^{|\mcs{k}|-2}e^{-(\alpha'_k-\alpha_k)(1-c)t}\right)}{\sum_{j''\in\mcs{k}}\bm{l}_{j''}^{(k)}+O\left(t^{|\mcs{k}|-2}e^{-(\alpha'_k-\alpha_k)(1-c)t}\right)}.
    \end{aligned}
\end{equation}
Finally, substituting into \eqref{eq:2star}, we can write
\begin{equation}
    \begin{aligned}
        \mathbb{P}(X(t)=j|X(t)\in\mcs{k},X(0)=i)&=\frac{1+\epsilon_t^{(j)}}{1+\epsilon'_t}\left(\frac{\bm{l}_j^{(k)}}{\sum_{j''\in\mcs{k}}\bm{l}_{j''}^{(k)}}+O\left(t^{|\mcs{k}|-2}e^{-(\alpha'_k-\alpha_k)(1-c)t}\right)\right)\\
        &=\frac{\bm{l}_j^{(k)}}{\sum_{j''\in\mcs{k}}\bm{l}_{j''}^{(k)}}+O\left(e^{-c't}+t^{|\mcs{k}|-2}e^{-(\alpha'_k-\alpha_k)(1-c)t}\right).
    \end{aligned}
\end{equation}
This completes the proof of our claim.

\end{appendices}

\bibliography{wolb-bibliography}% common bib file


\end{document}